\documentclass[10pt,twocolumn,letterpaper]{article}
\usepackage{iccv}
\usepackage{times}
\usepackage{epsfig}
\usepackage{graphicx}
\usepackage{amsmath}
\usepackage{amssymb}
\usepackage{booktabs}
\usepackage{tikz}
\usepackage{subcaption}
\usepackage{caption}
\usepackage{mathtools}
\usepackage{float}
\usetikzlibrary{positioning, shapes.geometric}
\usepackage[accsupp]{axessibility}
\usepackage[utf8]{inputenc} 
\usepackage[T1]{fontenc}   
\usepackage{CJKutf8}      
\usepackage{booktabs}
\usepackage{url}          
\usepackage{booktabs}     
\usepackage{amsfonts}   
\usepackage{nicefrac}   
\usepackage{microtype}  
\usepackage{xcolor}
\usepackage{multirow}
\usepackage{authblk}
\usepackage[accsupp]{axessibility}

\usepackage[pagebackref=true,breaklinks=true,letterpaper=true,colorlinks,linkcolor=red,bookmarks=false, citecolor=cyan]{hyperref}
\iccvfinalcopy

\begin{document}
\title{Improving Lens Flare Removal with General-Purpose Pipeline \\ and Multiple Light Sources Recovery}
%\vspace{-20cm}
\author[1]{Yuyan Zhou}
\author[1]{Dong Liang \thanks{Corresponding author: liangdong@nuaa.edu.cn}}
\author[1]{Songcan Chen}
\author[1]{Sheng-Jun Huang}
\author[2]{Shuo Yang}
\author[3]{Chongyi Li}
\affil[1]{MIIT Key Laboratory of Pattern Analysis and Machine Intelligence, College of Computer Science and Technology, 

Nanjing University of Aeronautics and Astronautics, Nanjing, China}
\affil[2]{Imaging Technology Group, DJI Innovations Co. Ltd., Shanghai, China}
\affil[3]{School of Computer Science, Nankai University, Tianjin, China 
\authorcr \{yuyanzhou, liangdong, s.chen, huangsj\}@nuaa.edu.cn, shuo.yang2@dji.com, lichongyi@nankai.edu.cn}
\renewcommand*{\Affilfont}{\normalsize}
\maketitle
\begin{abstract}
When taking images against strong light sources, the resulting images often contain heterogeneous flare artifacts. These artifacts  can importantly affect image visual quality and downstream computer vision tasks. While collecting real data pairs of flare-corrupted/flare-free images for training flare removal models is challenging, current methods utilize the direct-add approach to synthesize  data. 
However, these methods do not consider automatic exposure and tone mapping in image signal processing pipeline (ISP), leading to the limited generalization capability of deep models training using such data. Besides, existing methods struggle to handle multiple light sources due to the different sizes, shapes and illuminance of various light sources. In this paper, we propose a solution to improve the performance of lens flare removal by revisiting the ISP and remodeling the principle of automatic exposure in the synthesis pipeline and design a more reliable light sources recovery strategy.
The new pipeline approaches realistic imaging by discriminating the local and global illumination through convex combination, avoiding global illumination shifting and local over-saturation.
Our strategy for recovering multiple light sources convexly averages the input and output of the neural network based on illuminance levels, thereby avoiding the need for a hard threshold in identifying light sources. We also contribute a new flare removal testing dataset containing the  flare-corrupted images captured by ten types of consumer electronics. The dataset facilitates the verification of the generalization capability of flare removal methods. Extensive experiments show that our solution can effectively improve the performance of lens flare removal and push the frontier toward more general situations.
\end{abstract}

\section{Introduction}

\label{intro}
Lens flare artifacts commonly appear in the forms of halos, streaks, saturated blobs, and color bleeding~\cite{wu}. 
These artifacts can be roughly classified into two groups: scattering flare and reflective flare. 
Scattering flare occurs  due to  dust or wears in front of the lens, while reflective flare is caused by light reflection within the lens system. 
Physically, anti-reflection coating inside the lens system can partially suppress flare. 
However, in smartphone imaging with a simplified lens system and easily contaminated lens surfaces, flare is exacerbated.
Lens flare not only affects the visual quality of images  but also degrades the  performance of downstream computer vision tasks such as object detection in an automatic driving system. 
Removing lens flare from an image is an extremely challenging task since it is closely related to the properties of the light sources, such as the incident angle, location, size, intensity, and spectrum, as well as the heterogeneous lens types.

\begin{figure*}[!t]
\begin{center}
  \begin{subfigure}{0.23\textwidth}
  \centering
    \includegraphics[width=3.8cm,height=3.8cm]{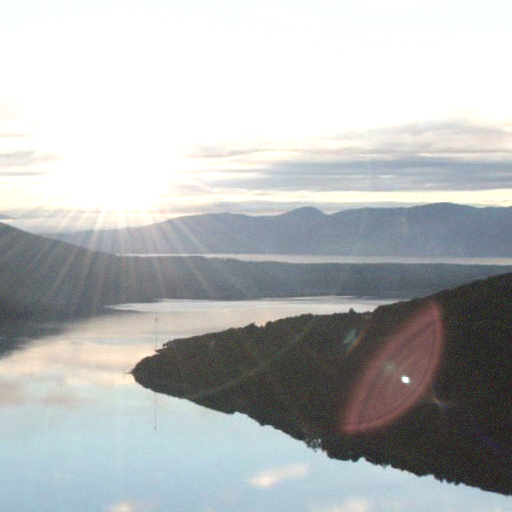}
    %\vspace{-.3cm}
    \caption{\small Direct-add \cite{wu,dai}}
  \end{subfigure}
  \quad
  \begin{subfigure}{0.23\textwidth}
    \centering
    \includegraphics[width=3.8cm,height=3.8cm]{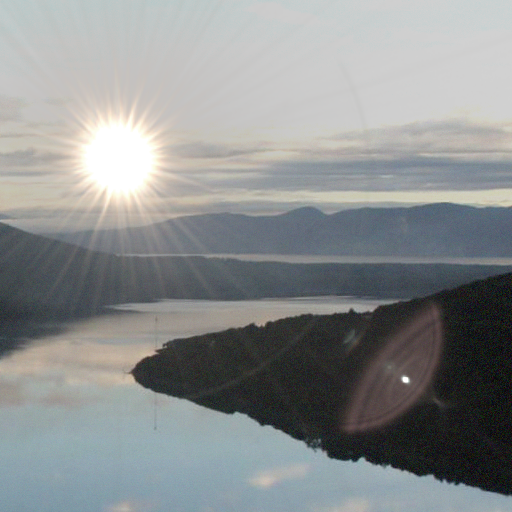}
        %\vspace{-.3cm}
    \caption{Our synthesis pipeline}
  \end{subfigure} 
  \quad 
  \begin{subfigure}{0.23\textwidth}
  \centering
    \includegraphics[width=3.8cm,height=3.8cm]{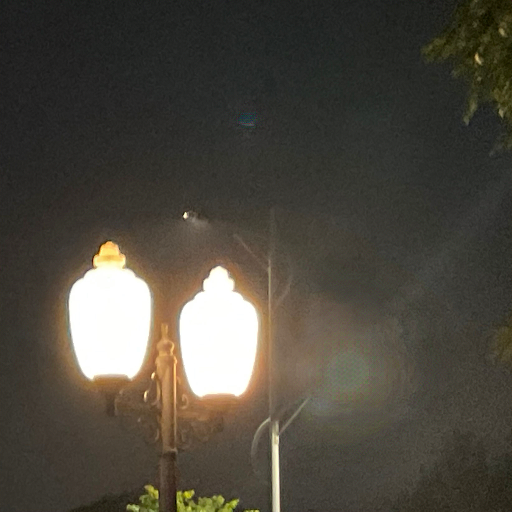}
    \caption{\small Recovered by~\cite{wu, dai}}
  \end{subfigure}
  \quad 
  \begin{subfigure}{0.23\textwidth}
    \centering
    \includegraphics[width=3.8cm,height=3.8cm]{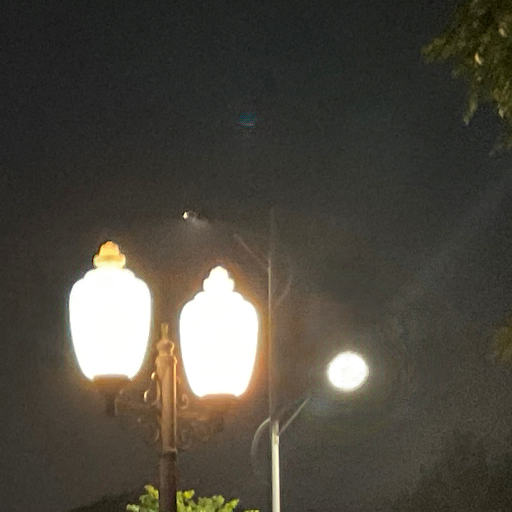}
    \caption{\small Recovered by Ours}
  \end{subfigure}
\end{center}
  \caption{Comparison of the flare-corrupted image and light source recovery in previous works and our method. Our method can synthesize a more realistic flare-corrupted image and preserve more natural light sources.
  }
  \label{1synthetic}
\end{figure*}

Like other low-level computer vision tasks such as reflection removal \cite{fan, li2020single}, low light enhancement~\cite{zerodce, scl, jin2022unsupervised}, and haze removal~\cite{liu,guo2022image, jin2023enhancing}, the lack of paired training data is the biggest obstacle in the task of flare removal. 
Creating large amounts of paired training data is time-consuming and labour-intensive.
To solve this issue, a recent work~\cite{wu} created a flare dataset with 2001 captured flare-only images and 3000 simulated flare-only images. 
To address the issue of trained models performing poorly in nighttime, a new dataset Flare7K~\cite{dai} was created specifically to remove nighttime flares.
However, these works assume that the flare-free and flare-only images are two independent layers and directly adds them in the RAW space. 
As the RAW formats of both flare and scene are not available, this work regards the inverse gamma transformed image as the RAW image, ignoring the typical tone mapping operator (TMO) in an image signal processing pipeline (ISP) (see Figure.~\ref{pipeline}).
Since the transformation from RAW to RGB image is irreversible, directly adding the two layers may suffer from the over-saturation issue with low contrast, as shown in Figure.~\ref{1synthetic}(a). 
Furthermore, most consumer cameras are equipped with auto-exposure (AE), which automatically adjusts the aperture and shutter speed to control the amount of light. 
Consequently, directly adding a flare image can brighten the scene, which is inconsistent with AE and causes an overall intensity distribution shift. (see Figure.~\ref{ds})

In addition to the drawbacks of the existing flare synthesis pipeline, the light source recovery problem still challenges current flare removal methods, particularly in recovering multiple light sources.
Most networks typically remove the light source along with the flare, as they cannot identify and separate the light source from the flare. 
To alleviate this problem, recent methods~\cite{wu,dai} tend to find the brightest connectivity component and apply a smoothing post-processing operation.
The failure to do so may result in an unrealistic light source appearance, as the failure case in Figure.~\ref{1synthetic}(c).

Unlike the previous works that focus on data preparation in daytime~\cite{wu} and nighttime~\cite{wu,dai} or design specific networks~\cite{qiao}, 
we provide two key insights to improve the performance of lens flare removal, both of which are ignored by the previous research:
(1) How to synthesize more realistic flare-corrupted images to simulate the general AE mode and takes tone mapping into consideration?  
(2) How to recover one or multiple light sources naturally and avoid the hard threshold?

To achieve that, we first revisit the ISP and remodel the optical synthesis principle.
Then we propose a solution to generate more realistic flare-corrupted images and preserve multiple light sources well in the final results. 
Rather than directly adding the scene and flare, our data synthesis pipeline generates flare-corrupted images by pixel-wise convex combinations between the scene and flare image in inverse gamma space. 
Our new pipeline effectively avoids the issues of global illumination shifting and local over-saturation in synthetic images.
Unlike previous methods, where the light sources are always affected along with the flare, our method can recover multiple light sources well. It convexly averages of the input and output of the neural network based on illuminance levels and avoids the hard threshold when identifying light sources.
In addition, we contribute a new flare removal testing dataset containing the flare-corrupted images captured by ten types of consumer
electronics to supplement existing lens flare datasets.
Extensive experiments demonstrate the effectiveness and contributions of our key designs.
Our main contributions are summarized below.
\begin{itemize}
  \item We systematically analyze the drawbacks of existing lens flare synthesis and creatively propose a new pipeline to generate more realistic flare-corrupted images and avoid illumination distribution shift for flare removal.
  \item We solve the challenging light source preservation issue in flare removal using an elegant strategy that can recover multiple light sources with heterogeneous shapes, illumination, and quantities.
  \item We contribute a new dataset that contains real flare-corrupted images captured by diverse consumer electronics, which provides an avenue to examine the generalization performance of flare removal methods.
\end{itemize}

\section{Related Work} 

\noindent \textbf{Physical Flare Removal.}
\label{Hard}
The most common optical solution to avoid lens flare is to apply an anti-reflection coating to the surface of lenses~\cite{antireflection}. 
It can weaken the reflection of light and greatly enhances the transmission in the lens system by utilizing destructive interference. 
However, it cannot completely reduce reflection, and particularly fails in the case, in which the light source is extremely bright.

\noindent \textbf{Computational Flare Removal.} 
Due to the complexity and diversity of the optical mechanisms, effective computational solutions for flare removal are rare. 
Traditional methods~\cite{traditional, traditional2, traditional3} can be separated into two steps: flare detection and removal. 
These methods detect flares  based on the strong assumptions on flares' illuminance, shape, and positions, and then use exemplar patches to inpaint the region. 
However, these methods can only remove partial flares as flares have various types and appearances.
Current deep learning-based de-flare methods are also scarce. 
Wu et al.~\cite{wu} directly added a flare image to a scene image to synthesize a flare-corrupted image to train a neural network. 
Qiao et al.~\cite{qiao} proposed a network trained on unpaired flare data, composed of a light source detection module, flare detection and removal, and generation module. 

\noindent \textbf{Lens Flare Dataset.} 
The main challenge in flare removal is the lack of paired training data. 
Wu et al.~\cite{wu} first proposed a semi-synthetic dataset containing 2001 captured and 3000 simulated flare images. 
To solve the limitations of Wu's dataset such as the limited lens flare type, especially in the nighttime, Dai et al.~\cite{dai} provided a synthetic dataset with diverse flare types, named Flare7K. 
Flare7K offers 5,000 scattering and 2,000 reflective flare images and consists of 25 types of scattering and 10 types of reflective flares.

\noindent \textbf{Computational Image Distortion.} 
Some recent works apply computational and learning-based approaches to 
reflection removal~\cite{reflection1, reflection2, reflection3}, rain removal~\cite{rainremoval1, rainremoval2, rainremoval3}, and
haze removal~\cite{hazeremoval1, hazeremoval2, hazeremoval3, hazeremoval4}. 
These methods attempt to decompose an image into original and corrupted components by training a neural network with specific training data. 
\section{Preliminaries}
\label{Preliminaries}
\subsection{Revisiting Image Signal Processing (ISP)}
\label{revisit}
\begin{figure}
    \centering
    \includegraphics[width=.42\textwidth]{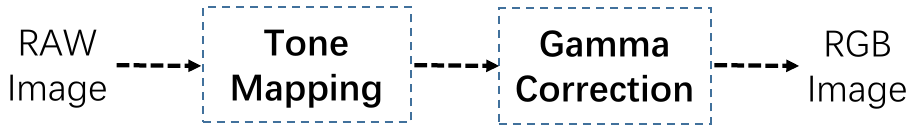}
    \includegraphics[width=.4\textwidth]{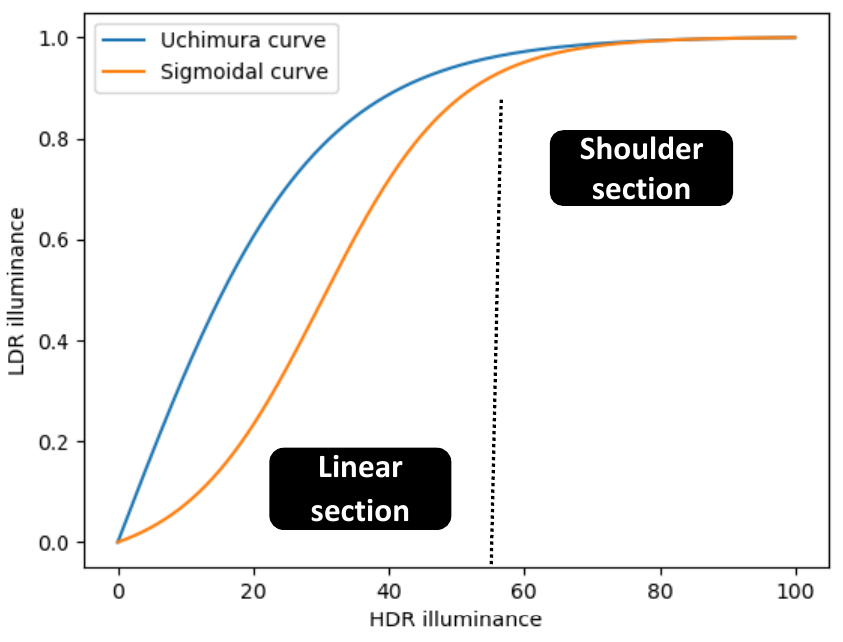}
\caption{The upper is a simplified image signal processing pipeline. The lower are two tone mapping operators, Uchimura and Sigmoid curve. Each camera has its specific tone mapping curve.}
  %\vspace{-0.5cm}
\label{pipeline}
\end{figure}

Photons received by sensors are transformed from analog signals to digital signals. 
The dynamic range of our ordinary life is in the range of [0, $10^6$]~\cite{tmoreview}; however, human visual system (HVS)
can perceive a range of [0, 1.6$\times10^4$]. 
The direct linear transformation can lead to image detail loss and substantial contrast reduction. 
Since HVS is more sensitive to contrast rather than absolute illuminance, a nonlinear function called tone mapping operator (TMO) was designed to map the illuminance in the domain $[0,+\infty)$ (High Dynamic Range (HDR)) to the output ranged in $[0,1]$ (Low Dynamic Range (LDR)), which can preserve image contrast.  
As shown in Figure.~\ref{pipeline}, the section in TMO that maps larger illuminant values in HDR to 1 in LDR asymptotically is called the \textbf{Shoulder section}. 
Before the shoulder section, the \textbf{Linear section} is the most linear portion and controls the mid-tones scale of the image. 
Different digital cameras use different tone-mapping operators. 
When the tone mapping operator is always inreversible and not offered, it is difficult to recover the RAW image from the RGB image. 

As shown in Figure.~\ref{pipeline}, after tone mapping, ISP applies a gamma correction to fit HSV further. 
Gamma correction is also a non-linear operation used to encode luminance values in image display systems. 
It is typically defined by a simple power-law expression. It optimizes the illuminance when encoding an image, by taking advantage of the non-linear manner in which humans perceive illuminance and color.

\subsection{Analyzing Flare Image Synthesis with ISP}
% \vspace{-.1cm}
Current methods~\cite{dai, wu} synthesize paired data for flare removal based on a critical observation that lens flare is an \textbf{additive layer} on the underlying image in RAW space. 
Obviously, this assumption is invalid in RGB space and will cause overflow. 
To this end, current methods~\cite{dai, wu} regard the gamma-inversed image of RGB image
%(using a power-law function)
as a RAW image and directly add a flare-free and a flare-only image in the gamma-inversed space to synthesize a flare-corrupted image, which can be expressed as
\begin{equation}
  I=S+F+N(0, \sigma^2),
   \label{eq:da}
\end{equation}
where $S$ is a flare-free gamma-inversed image, $F$ is a flare-only  gamma-inversed image and $N(0, \sigma^2)$ denotes random Gaussian noise used to narrow domain gap. 

We argue that using a gamma-inversed image as its RAW image is unreasonable. 
As introduced in Sec. \ref{revisit}, the RAW image is the first tone mapped from  HDR to LDR by the TMO. 
Then the LDR image is gamma-corrected to be the final image. 
Adding two RGB images in RAW space needs its TMO $T$ and its inverse function like
\begin{equation}
  I=T(T^{-1}(S)+T^{-1}(F))+N(0, \sigma^2).
   \label{true}
\end{equation}
Since the camera-specific TMO is irreversible and unavailable, the current methods regard the tone mapping of scene and flare image as \textbf{linear identity mapping}, and the gamma-inversed image is treated as a RAW image.
For the flare-free scene, most pixels are in the \textbf{Linear section} of TMO. 
Treating tone mapping as linear identity mapping is reasonable. 
Nevertheless, regarding the flare image, many pixels around the light source are in the \textbf{Shoulder section} of TMO. 
Hence, the TMO of flare images cannot be treated as linear identity mapping. 
Therefore, the range of both contrast and color  near the light sources in the image synthesized by this method would be flattened, as demonstrated in Figure.~\ref{1synthetic}(a).   
\subsection{Rethinking More Reasonable Solution}
\label{rethinking}
So how can we add two layer in RAW space only using RGB image? 
With this question, like Brooks et al.~\cite{unprocessing}, we assume HDR domain is $[0,1]$ and use the smooth step TMO $T(x) = 3x^2-2x^3$ for analyzing.
Since pixels in the flare-only image range from the brightest to the darkest part, we denote pixels in the two parts in RGB space $b_{ij}$ and $d_{pq}$. 

Specifically, we first focus on the brightest part when adding a scene layer pixel $s_{ij}$ which is in the linear section of TMO in RAW space.
Given $\frac{T^{-1}(s_{ij})}{T^{-1}(b_{ij})}=\epsilon_1$, where $\epsilon_1$ is a small quantity. First, representing $b_{ij}$ using $T^{-1}(b_{ij})$: %\lichongyi{what is $b_{ij}$? what is the mean of `evaluating'?}
\begin{align}
    T(T^{-1}(b_{ij})) &= 3T^{-1}(b_{ij})^2-2T^{-1}(b_{ij})^3 \\
    &\approx 3T^{-1}(b_{ij})^2-2T^{-1}(b_{ij})^2\\
    &=T^{-1}(b_{ij})^2.
\end{align}
%\lichongyi{what is T?}
Because $T^{-1}(b_{ij})$ tends to 1,  $T^{-1}(b_{ij})^3\approx T^{-1}(b_{ij})^2$. Then representing $s_{ij}$ using $T^{-1}(b_{ij})$:
%So direct add two pixels in the RAW space, and then tone mapping can be expressed as below.     
\begin{align}
    T(T^{-1}(s_{ij})) &= T(\epsilon_1 T^{-1}(b_{ij})) \\
    &= 3\epsilon_1^2T^{-1}(b_{ij})^2-2\epsilon_1^3T^{-1}(b_{ij})^3\\
    &\approx 3\epsilon_1^2T^{-1}(b_{ij})^2.
\end{align}
Since $\epsilon_1^3$ is an infinitesimal of a higher order than $\epsilon_1^2$, it can be ignored. Now we can represent Eq.~\eqref{true} using RGB image value $b_{ij}$ and $s_{ij}$: %\lichongyi{cannot understand why use evaluation and the purpose of these equations.}
\begin{align}
    &T(T^{-1}(b_{ij})+T^{-1}(s_{ij}))\\
    %=& T((1+\epsilon_1)T^{-1}(b_{ij})) \\
    =& 3(1+\epsilon_1)^2T^{-1}(b_{ij})^2-2(1+\epsilon_1)^3T^{-1}(b_{ij})^3 \\
    \approx& 3(1+\epsilon_1)^3T^{-1}(b_{ij})^2-2(1+\epsilon_1)^3T^{-1}(b_{ij})^2\\
    %=& (1+\epsilon_1)^3T(T^{-1}(b_{ij}))\\
    %=&(1+3\epsilon_1+3\epsilon_1^2)T(T^{-1}(b_{ij})) + \frac{\epsilon_1}{3}T(T^{-1}(s_{ij})).\\
    =& (1+3\epsilon_1+3\epsilon_1^2)b_{ij} +  \frac{\epsilon_1}{3}s_{ij}
\end{align}
Here the final result lies in the range of $[0, 1+3\epsilon_1+3\epsilon_1^2+\frac{\epsilon_1}{3}]$. Since $1+3\epsilon_1+3\epsilon_1^2+\frac{\epsilon_1}{3}\approx 1$, we use it to divide the final result and obtian
\begin{align}
    &T(T^{-1}(b_{ij})+T^{-1}(s_{ij}))\\
    & \approx \frac{1+3\epsilon_1+3\epsilon_1^2}{1+3\epsilon_1+3\epsilon_1^2+\frac{\epsilon_1}{3}}b_{ij} + \frac{\frac{\epsilon_1}{3}}{1+3\epsilon_1+3\epsilon_1^2+\frac{\epsilon_1}{3}}s_{ij}\\
    &\coloneqq (1-\epsilon_2)b_{ij}+\epsilon_2s_{ij},
\end{align}
where $\epsilon_2$ is used to denote the coefficient of the second term.
When $\epsilon_1$ tends to 0, the weight of the first term $(1-\epsilon_2)$ tends to 1, and the weight of the second term $\epsilon_2$ tends to 0. 
Eq. $(10)$ shows a daily observation that when a strong light source appears in an image, all image details will be severely occluded.
Therefore, the weight of the scene image is very small, and the weight of the light source tends to be 1. 
It is worth noting that not all scene image pixels are in the linear section, sometimes with highlights. 
In this case, when adding the two saturated pixels, it will be clipped and tone-mapped to be 1. 
Eq. $(17)$ used in our method also leads to 1, which is consistent with the real case.

For the darkest part, pixels $d_{pq}$ are tend to 0, let $\epsilon_3 = \frac{T^{-1}(d_{pq})}{T^{-1}(s_{pq})}$. First, we represent $s_{pq}$ and $d_{pq}$ using $T^{-1}(s_{pq})$:
\begin{align}
    T(T^{-1}(s_{pq})) &= 3T^{-1}(s_{pq})^2 - 2T^{-1}(s_{pq})^3\\
    T(T^{-1}(d_{pq})) &= 3\epsilon_3^2 T^{-1}(s_{pq})^2 - 2\epsilon_3^3T^{-1}(s_{pq})^3\\
                      &\approx 3\epsilon_3^2T^{-1}(s_{pq})^2
\end{align}
According to Wu et al.~\cite{wu}, to synthesize the flare-corrupted image, we need to add flare image and scene image in pre-tonemapping space and then  map them to the RGB space:
\begin{align}
    &T(T^{-1}(s_{pq})+T^{-1}(d_{pq})) \\
    =& 3(1+\epsilon_3)^2T^{-1}(s_{pq})^2 - 2(1+\epsilon_3)^3T^{-1}(s_{pq})^3  \\
    \approx & 3(1+\epsilon_3)^3T^{-1}(s_{pq})^2- 2(1+\epsilon_3)^3T^{-1}(s_{pq})^3\\
    =& (1+3\epsilon_3+3\epsilon_3^2)s_{pq} + \frac{\epsilon_3}{3}d_{pq}
\end{align}
%This \lichongyi{this means what?} is the same as Eq. $(9)$ \lichongyi{Eq. 9 in the main paper?}. 
Divide it using $1+3\epsilon_3+3\epsilon_3^2+\frac{\epsilon_3}{3}\approx 1$ and denote the coefficient of the second term $\epsilon_4$, we have
\begin{equation}
    T(T^{-1}(s_{pq})+T^{-1}(d_{pq})) = (1-\epsilon_4)s_{pq} + \epsilon_4d_{pq}
\end{equation}
This shows that the darkest part of the flare layer hardly influences the final image. 
Its weight $\epsilon_4\approx 0$, so it plays a negligible role in the final image.

We can see that when the pixels in the scene image add with flare image pixels from brightest to darkest, the weight of the scene image becomes more significant from $\epsilon$ to $1-\epsilon$, and the weight of the flare image becomes smaller from $1-\epsilon$ to $\epsilon$. 
Thus, when we blend two images, we perform a convex combination for each pixel. 
Concretely, if the pixel in the flare image is bright, it will be assigned a larger weight. Otherwise, %if the pixel in the flare image is dark,
it will be assigned a smaller weight. 
%

 %\vspace{-.2cm}
\section{Proposed Flare-Corrupted Image Generation}
 %\vspace{-.2cm}
\subsection{Our Pipeline}
 %\vspace{-.1cm}
\label{syn}
%Our flare-corrupted image generation approach is easy to implement.
Motivated by the discussion in Sec. \ref{Preliminaries}, we assign weight to every pixel in the flare image and scene image in gamma-inversed space according to its illuminance via a convex combination. 
The process can be divided into the following three steps
(1) Calculate the illuminance matrix of the flare image.
(2) Assign a weight to every pixel according to the illuminance matrix.
(3) Blend the scene layer and flare layer by convex combination.
We detail the process below.

\noindent\textbf{Calculate illuminance matrix:}
Calculating the illuminance matrix $I_F$ of the flare layer can be achieved by adding its RGB channel and then normalizing it to $[0,1]$.
\begin{equation}
 I_F = \frac{1}{255\times 3}\sum\limits_{c=r,g,b}F_c.
  \label{eq:I}
\end{equation}

\noindent\textbf{Assign a weight to every pixel:}
We determine the weight of every pixel by the illuminance matrix $I_{F}$.
As discussed in Sec. \ref{rethinking}, if the pixel value of $I$ is larger, we assign the corresponding element of $W$ a larger weight. If the pixel value of $I$ is smaller, we assign the corresponding element of $W$ a smaller weight. We use a function $f$ to determine the weigh according to the illuminance,
\begin{equation}
  W= f(I_{F}).
\end{equation}
The weight function $f$ is similar to TMO and we use a simple sigmoid function as the weight function, expressed as:
\begin{equation}
  f(x) =  \frac{1}{1+e^{p(x-q)}},
  \label{weight function}
\end{equation}
where $q=0.5$ and $p$ is sampled from uniform distribution $U[4, 7]$.

\noindent\textbf{Blend the scene layer and flare layer by convex combination:}
Using the calculated weight matrix, we can blend the scene $S$ and flare layers $F$ by convex combination. 
Following previous methods~\cite{wu, dai}, we add the same Gaussian noise to narrow the domain gap, whose variance is sampled once per image from chi-square distribution $\sigma^2\sim 0.01 \chi^2 $.   
%
%At last, the flare-corrupted image can be expressed using scene image $S$ and flare image $F$:
\begin{equation}
  I = (1-W)\odot S + W\odot F + N(0, \sigma^2),
\end{equation}
where $\odot$ means element-wise multiplication.

\subsection{Rationality Analysis}%Interpretation with AE
\label{AE}
 \begin{figure}[t]
  \centering
  \centering
    \includegraphics[width=.49\textwidth]{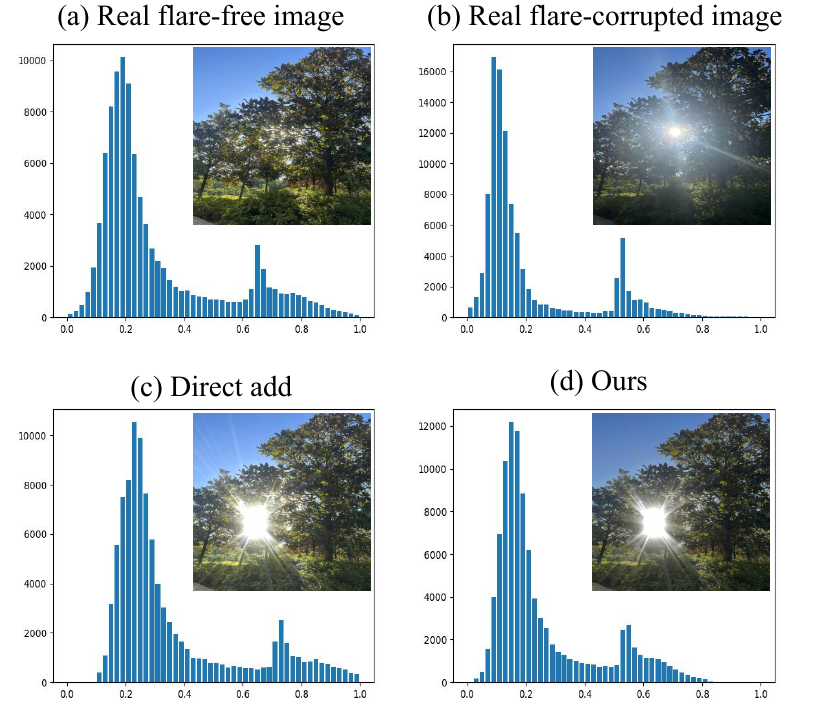}
 \caption{\small Intensity Distribution of a real flare-free image (a), a real flare-corrupted image (b), and synthetic flare-corrupted images (c-d).  The distribution of the image synthesized by our method aligns well with the real case. (X-axis: intensity value from 0 to 1, Y-axis: pixel intensity counting)
 }
 \label{ds}
\end{figure}

When a digital camera is pointed at a strong light source, i.e. back-lighting photography, the automatic exposure mode (AE) automatically adjusts the aperture setting and shutter speed to avoid overexposure. 
The faster shutter speed and smaller aperture size can reduce the light entering into the lens system, which darkens the background. 
To further verify our analysis that the shutter speed will be reduced, we used iPhone 13 pro to take 100 images with and without strong light sources and calculate the average shutter speed. 
The average shutter speed of images without and with a light source is $9.85\times 10^{-4}$s and $7.52\times 10^{-5}$s respectively.
We also present a set of visual examples in Figure.~\ref{ds}
, where the x-axis represents intensity values and the y-axis indicates the number of pixels.  
As shown, in the two real images, compared with the image without a strong light source, the intensity distribution of the image with a strong light source slightly moves to the darker part and the dynamic range of the distribution becomes narrower.
\begin{figure*}[htb]
  \begin{center}
\begin{subfigure}{0.137\linewidth}
      %\caption{input}
      \includegraphics[width=2.43cm,height=2.43cm]{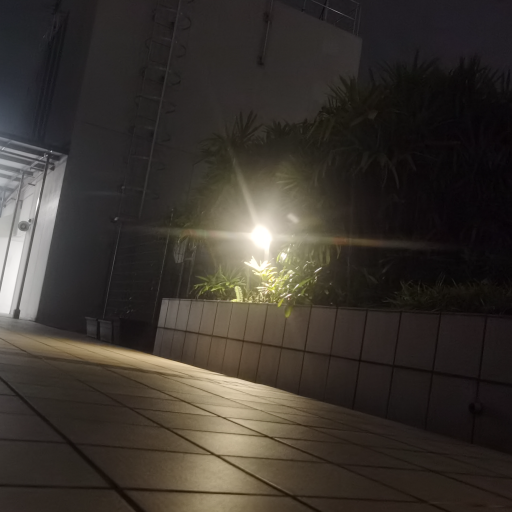}
    \end{subfigure}
    \begin{subfigure}{0.137\linewidth}
      %\caption{flare spot removal}
      \includegraphics[width=2.43cm,height=2.43cm]{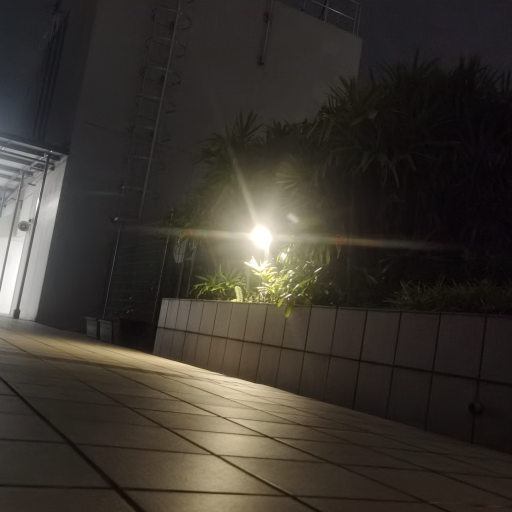}
    \end{subfigure}
    \begin{subfigure}{0.137\linewidth}
      %\caption{dereflection}
      \includegraphics[width=2.43cm,height=2.43cm]{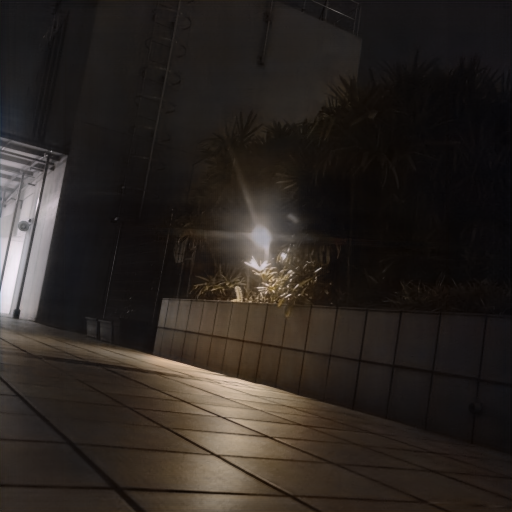}
    \end{subfigure}
    \begin{subfigure}{0.137\linewidth}
      %\caption{dehaze}
      \includegraphics[width=2.43cm,height=2.43cm]{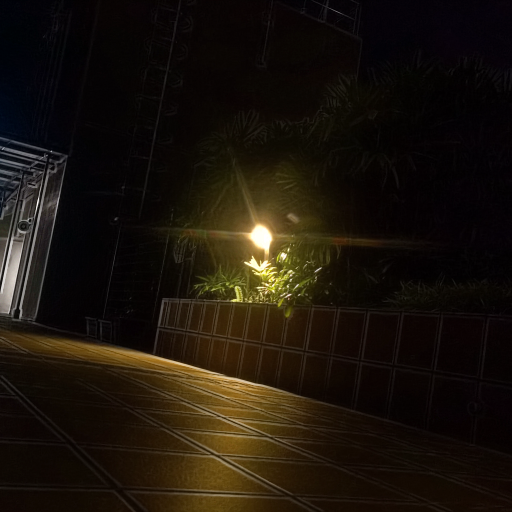}
    \end{subfigure}
    \begin{subfigure}{0.137\linewidth}
     % \caption{wu + wu's dataset}
      \includegraphics[width=2.43cm,height=2.43cm]{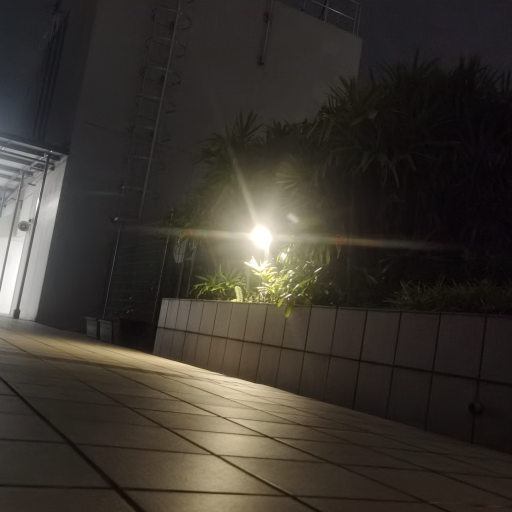}
    \end{subfigure}
    \begin{subfigure}{0.137\linewidth}
     % \caption{wu + wu's dataset}
      \includegraphics[width=2.43cm,height=2.43cm]{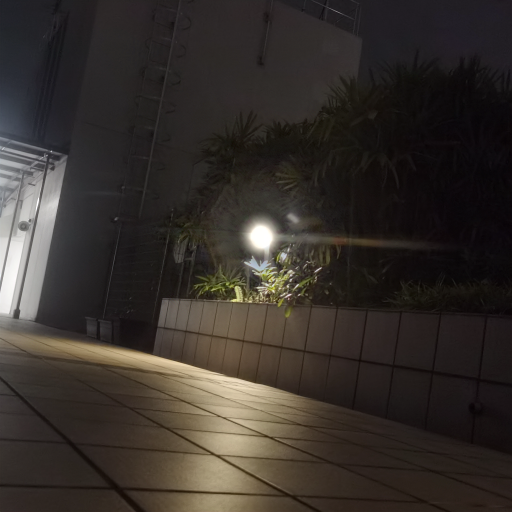}
    \end{subfigure}
    \begin{subfigure}{0.137\linewidth}
      %\caption{ours + wu's dataset}
      \includegraphics[width=2.43cm,height=2.43cm]{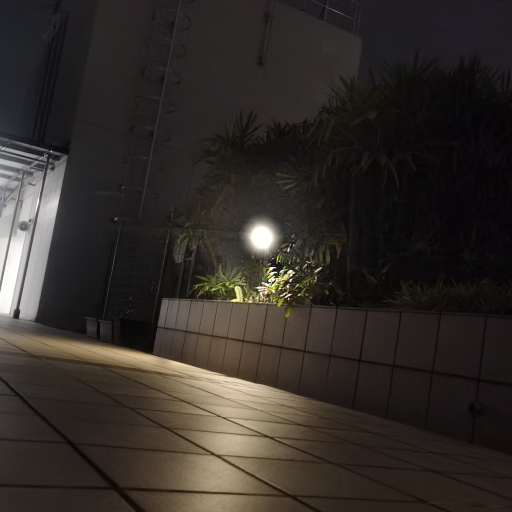}
\end{subfigure}\\ 
    \begin{subfigure}{0.137\linewidth}
      \includegraphics[width=2.43cm,height=2.43cm]{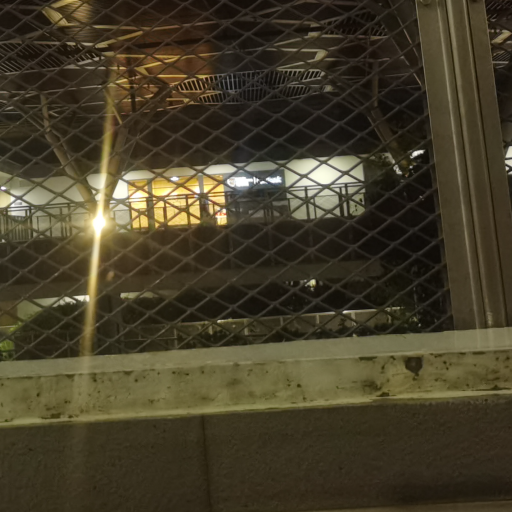}
    \end{subfigure}
    \begin{subfigure}{0.137\linewidth}
      \includegraphics[width=2.43cm,height=2.43cm]{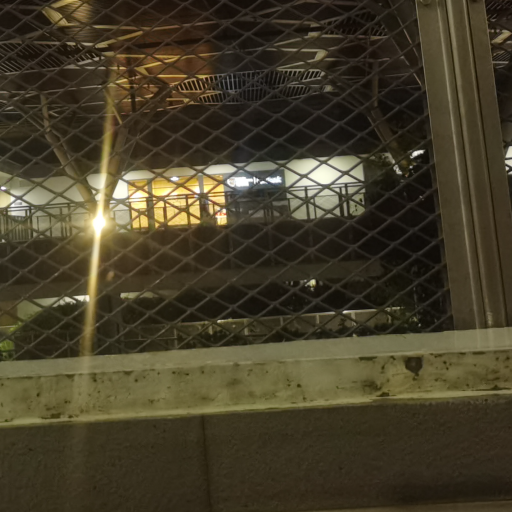}
    \end{subfigure}
    \begin{subfigure}{0.137\linewidth}
      \includegraphics[width=2.43cm,height=2.43cm]{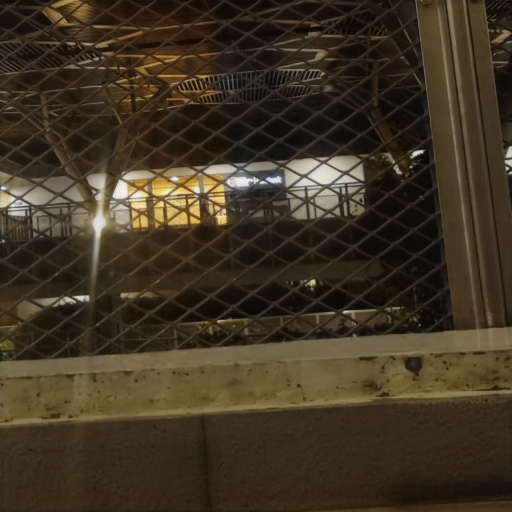}
    \end{subfigure}
    \begin{subfigure}{0.137\linewidth}
      \includegraphics[width=2.43cm,height=2.43cm]{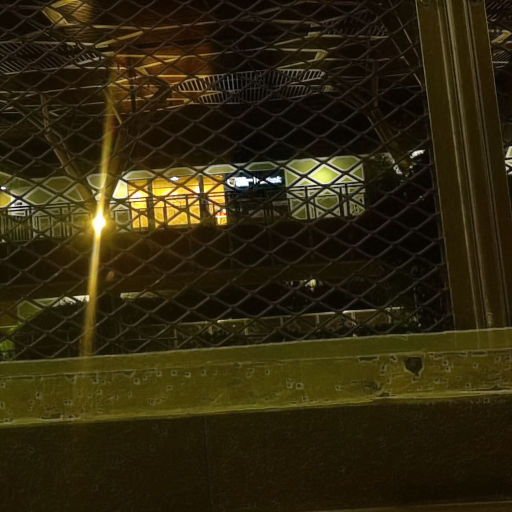}
    \end{subfigure}
    \begin{subfigure}{0.137\linewidth}
      \includegraphics[width=2.43cm,height=2.43cm]{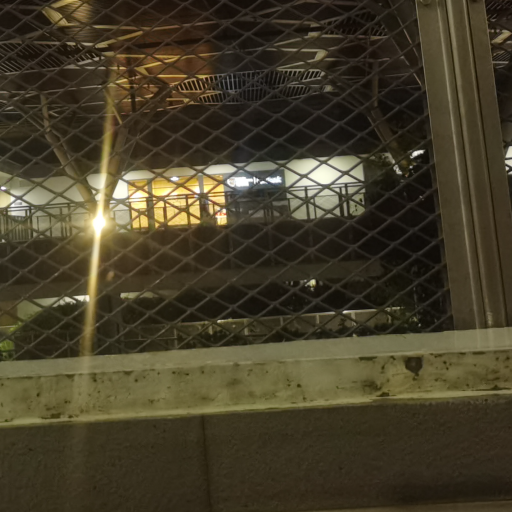}
    \end{subfigure}
    \begin{subfigure}{0.137\linewidth}
      \includegraphics[width=2.43cm,height=2.43cm]{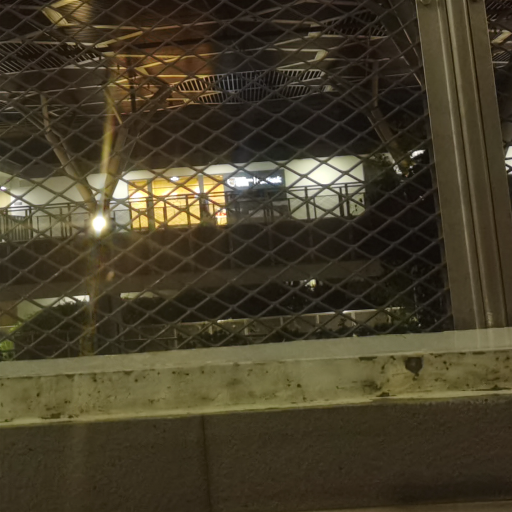}
    \end{subfigure}
    \begin{subfigure}{0.137\linewidth}
      \includegraphics[width=2.43cm,height=2.43cm]{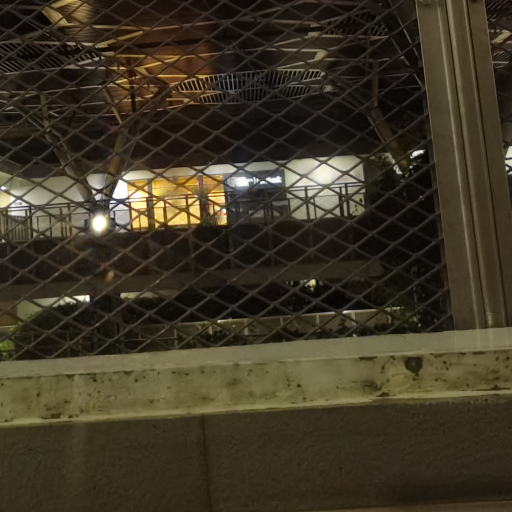}
\end{subfigure}\\  %%%%%%%%%%%%%%%%%%%%%%%%%%%%%%%%%%%%%%%%%%%%%%%%%
    \begin{subfigure}{0.137\linewidth}
      \includegraphics[width=2.43cm,height=2.43cm]{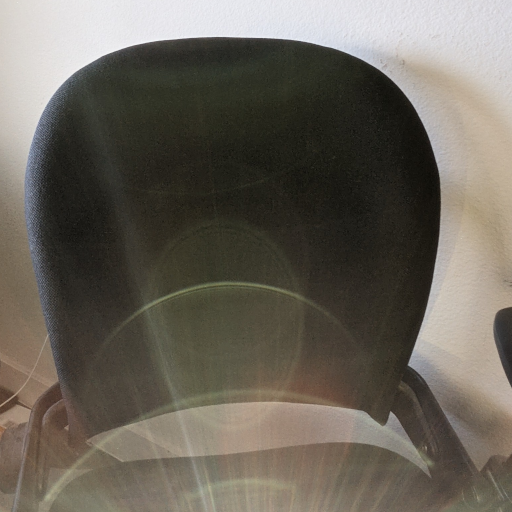}
      \caption{\footnotesize Input}
    \end{subfigure}
    \begin{subfigure}{0.137\linewidth}
      \includegraphics[width=2.43cm,height=2.43cm]{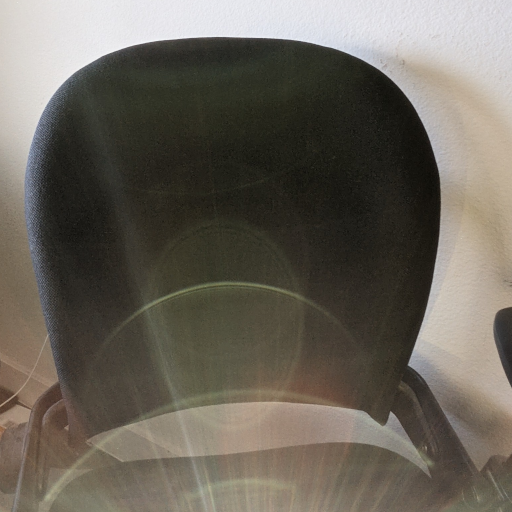}
      \caption{\footnotesize Asha et al.~\cite{traditional}}
    \end{subfigure}
    \begin{subfigure}{0.137\linewidth}
      \includegraphics[width=2.43cm,height=2.43cm]{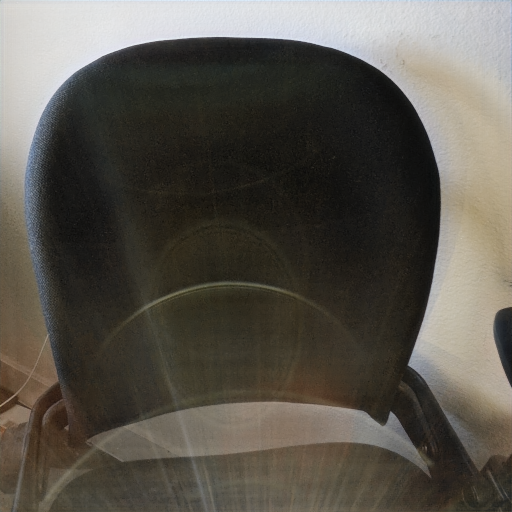}
      \caption{\footnotesize Zhang et al.~\cite{zhang}}
    \end{subfigure}
    \begin{subfigure}{0.137\linewidth}
      \includegraphics[width=2.43cm,height=2.43cm]{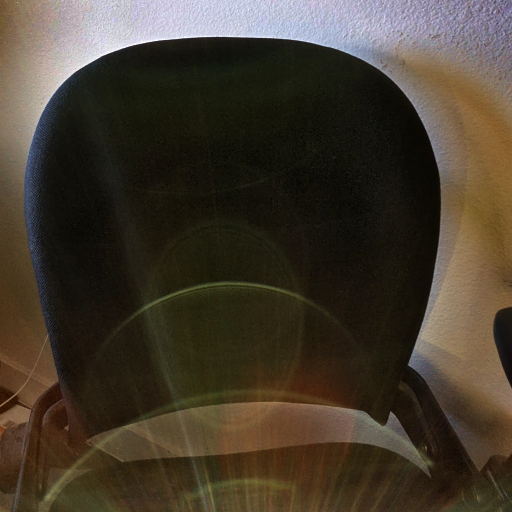}
      \caption{\footnotesize He et al.~\cite{he}}
    \end{subfigure}
    \begin{subfigure}{0.137\linewidth}
      \includegraphics[width=2.43cm,height=2.43cm]{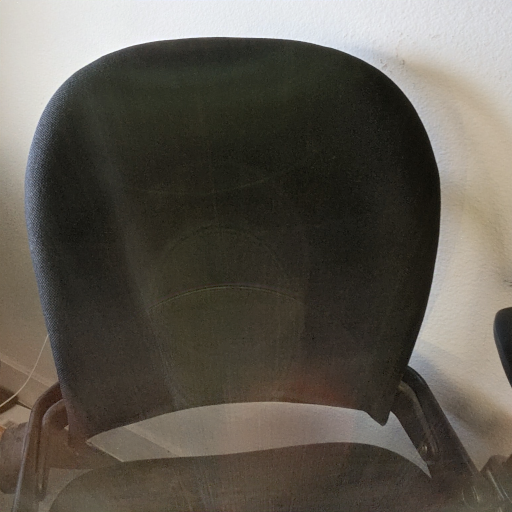}
      \caption{\footnotesize Wu et al.~\cite{wu}}
    \end{subfigure}
    \begin{subfigure}{0.137\linewidth}
      \includegraphics[width=2.43cm,height=2.43cm]{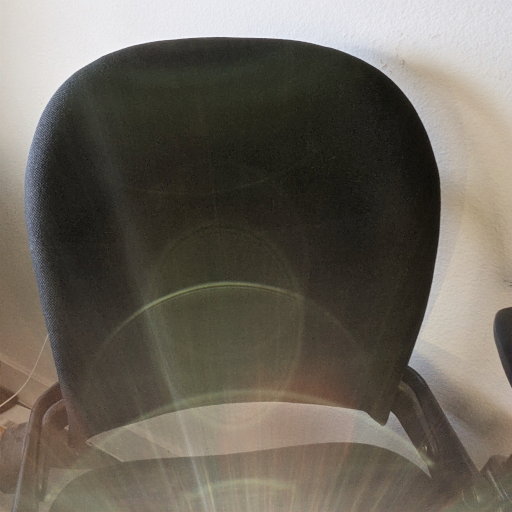}
      \caption{\footnotesize Dai et al.~\cite{dai}}
    \end{subfigure}
    \begin{subfigure}{0.137\linewidth}
      \includegraphics[width=2.43cm,height=2.43cm]{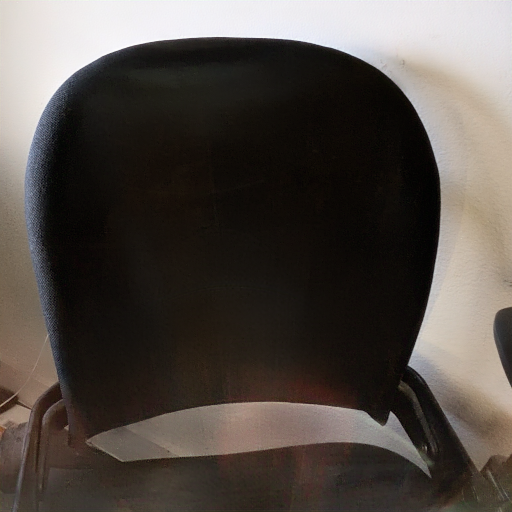}
      \caption{\footnotesize Ours}
    \end{subfigure}
  \end{center}
    \caption{\small Qualitative comparison on \cite{wu, dai} real test images with different method using U-Net~\cite{UNet}.}
    \label{5}
\end{figure*}
\begin{table*}\small
  \caption{\small Quantitative comparison with different methods on Flare7k test set.}
\setlength\tabcolsep{0.5pt}
  \centering
  \scalebox{0.79}{
  \begin{tabular}{ccccccccccc}
    \toprule  
    $\;$Method$\;$ & $\;$Dehaze \cite{he}$\;$ & $\;\;$Dereflection~\cite{zhang}$\;\;$ & $\;\;$Dai et al.~\cite{dai}$\;\;$ & $\;\;$Dai et al.~\cite{dai}$\;\;$ & $\;\;$Wu et al.~\cite{wu}$\;\;$ & $\;\;$Wu et al.~\cite{wu}$\;\;$ & $\;\;$Ours$\;\;$ & $\;\;$Ours$\;\;$ & $\;\;$Ours$\;\;$ & $\;\;$Ours$\;\;$\\
      \midrule
        Training set & $\times$ & pretrained  &  Flare7K~\cite{dai} &  Flare7K~\cite{dai} & Wu~\cite{wu} & Wu~\cite{wu} & Flare7K~\cite{dai} & Flare7K~\cite{dai} & Wu~\cite{wu} & Wu~\cite{wu} \\
        \midrule
    Model & $\times$ & CEILNet~\cite{cenet} & U-Net~\cite{UNet} & $\;\;$Uformer~\cite{Uformer}$\;\;$ & U-Net~\cite{UNet} & $\;\;$Uformer~\cite{Uformer}$\;\;$ & $\;\;$U-Net~\cite{UNet}$\;\;$ & $\;\;$Uformer~\cite{Uformer}$\;\;$ & $\;\;$U-Net~\cite{UNet}$\;\;$ & $\;\;$Uformer~\cite{Uformer}$\;\;$\\
    \midrule
    PSNR  & 19.7 & 23.3 & 25.4 & 25.7 & 23.6 & 23.7 & 25.3 & 25.7 & 25.9 & \textbf{26.3}\\
    SSIM~\cite{ssim} & 0.68 & 0.872 & 0.876 & 0.879 & 0.870 & 0.863 & 0.884 & 0.890 & \textbf{0.896} & 0.884 \\
    \bottomrule
  \end{tabular} }
  \label{tab}
  %\vspace{-.3cm}
\end{table*}
In the case of directly adding a scene and a flare image, the illuminance distribution in the synthetic image moves to the brighter illuminance part (as shown in the of Figure.~\ref{ds} (c)) 
The distribution shift~\cite{distribution} of training data will make the deep model biased to the training data and thus performs poorly in real cases.
In contrast, our method darkens the scene layer in the synthetic image, and the distribution of the scene layer moves to the darker part as shown in Figure.~\ref{ds}, which is consistent with the real case.
  
\section{Proposed Light Source Recovery}
\label{light}
Commonly, when a flare image is processed by a trained neural network, the light source in the image is treated as a flare and removed~\cite{wu}. 
However, the task of flare removal is to remove the flare and preserve the light source, we need to post-process the output of the neural network to recover light sources.
To address this issue, current methods~\cite{wu, dai} choose the brightest part of a flare image and set an illuminance threshold to determine whether it is a light source. 
As the illuminance threshold of the daytime light source differs from the nighttime light source~\cite{dai}, it is difficult to find a optimal threshold for both daytime and nighttime cases.

Based on the fact that the light source is always in the shoulder section of TMO, we use the same pipeline as mentioned in Sec.~\ref{syn}  for light source recovery. 
Specifically, we first choose a strong convex function with a larger second-order derivative as a weight function. 
The function can suppress the weight in the linear section and only assign larger weights to the pixels in the Shoulder section. 
We choose is  $x^{\alpha}$ as the weight function. 

Compared with the Sigmoid function in Eq.~\eqref{weight function} 
the strong convex function can ensure that only the light source in the original is blended into the final image. 
$\alpha$ determines what will be blended into the output of the neural network. 
When $\alpha \to +\infty$, the weight of the light source tends to one and the weight of other parts tends to 0, so only the light source will be recovered. When $\alpha \to -\infty$, the weight of the input image tends to 1, and the final image tends to be the input image, which means that both the light source and the flare will be recovered (See Sec.~\ref{recover} for further analysis). Thus, we choose $\alpha=15$ as default setting to recover light sources. The process pipeline can be expressed as
\begin{equation}
  I_{\text{input}} = \sum\limits_{c=r,g,b}C_c,
\end{equation}
\begin{equation}
  W_r = (\frac{I_{\text{input}}-\min I_{\text{input}} }{\max I_{\text{input}} - \min I_{\text{input}} })^{\alpha},
\end{equation}
\begin{equation}
  I_{\text{final}} = (1-W_r)\odot N(C) + W_r\odot C,
\end{equation}
\begin{table*}[htb]
\centering
\caption{User study. The result is similar to quantitative evaluation. There are 2001 images in~\cite{wu}'s flare dataset is captured in real life, while Flare7K \cite{dai} dataset is all synthetic. Our method trained using Wu et al.~\cite{wu} dataset performs better.}
% \vspace{-.1cm}
	 %\scalebox{0.8}{
	\begin{tabular}{lccc|ccc}
		\toprule 
		& \multicolumn{3}{c}{Trained on \cite{wu} dataset} & \multicolumn{3}{c}{Trained on \cite{dai} dataset} \\
		\hline
		Test dataset & \cite{wu} dataset & \cite{dai} dataset & Our dataset & \cite{wu} dataset & \cite{dai} dataset & Our dataset\\
		\hline
		Ours: Deflarespot\cite{traditional} & $\;$100\%:$\;\,$0\% & $\;$ 100\%:$\;\,$0\% & $\;\;\,$93\%:$\;\,$7\% & $\;\;\,$90\%:10\% & $\;$ 100\%:$\;\,$0\% & $\;\;\,$95\%:$\;\,$5\% \\
		Ours: Dehaze\cite{he} & $\;\;\,$90\%:10\% & $\;\;\,\,$93\%:$\;\,$7\%  & $\;$100\%:$\;\,$0\% & $\;\;\,$72\%:28\% & $\;\;\,\,$93\%:$\;\,$7\%  & $\;\;\;\,$87\%:13\%\\
		Ours: Dereflection\cite{zhang} & $\;\;\,$95\%:$\;\,$5\% & $\;\;\;\,$79\%:21\%  & $\;\;\;\,$87\%:13\% & $\;\;\,$51\%:49\% & $\;\;\,\,$95\%:$\;\,$5\%  & $\;\;\;\,$76\%:24\%\\
		Ours: Wu\cite{wu} & $\;\;\,$55\%:45\% & $\;$100\%:$\;\,$0\% & $\;$100\%:$\;\,$0\% & $\;\;\,$52\%:48\% & $\;\;\;\,$57\%:43\% & $\;\;\;\,$54\%:46\%\\
		\hline
	\end{tabular}%
	%}
  %\vspace{-.6cm}
	\label{user}
\end{table*}%
where $C$ denotes the input real flare-corrupted image, $I_{\text{input}}$ denotes the illuminance matrix of $C$, $W_r$ denotes the weight matrix used for light source recovery, $N(C)$ denotes the output of the neural network, and $I_{\text{final}}$ is the light source recovered flare-free image we desired. Note that that we use min-max normalization in Eq.\eqref{eq:I} instead of dividing by $255\times 3$.
Such operation guarantees that the weight of the brightest part in the input image will always be assigned to 1, i.e., the light source can be recovered.
%  \vspace{-.1cm}
\section{Experiments}
\subsection{Flare Removal Comparison}
We compare the results of our method with the traditional flare removal method~\cite{traditional} and deep models \cite{wu, dai}'s approach. Since the flare is also introduced by reflection and dust between and in front of the lens, we also compare our method with reflection removal~\cite{zhang} and haze removal~\cite{he}. 
Since Wu et al.’s work~\cite{wu} is most related to our work, we follow it to use a U-Net~\cite{UNet} as our flare removal baseline network. 
We also test a recently proposed transformer-based U-Net: Uformer~\cite{Uformer}. 
Both Wu et al.~\cite{wu} and Dai et al.~\cite{dai} use their flare datasets and apply the direct-add algorithm on the clear image dataset provided in [26] to synthesize flare-corrupted images. For fair comparison, we separately use Dai et al.~\cite{dai} and Wu et al.\cite{wu}'s flare datasets and the same clean image dataset. Differently, we apply our proposed pipeline to synthesize flare-corrupted images and compare our results with them. We implement our method with Tensorflow on a NVIDIA GTX 3090 GPU.  We also provide the implementation by MindSpore. 

\noindent\textbf{Qualitative Evaluation: } 
Figrue.~\ref{5} shows the visual comparison of different methods. 
As we can see in the second column, the traditional flare removal method~\cite{traditional} cannot remove scattering and reflection flare with different shapes. 
The third and fourth columns show that de-reflection~\cite{zhang} and dehaze~\cite{he} exhibit some ability to remove lens flare but cannot remove flare thoroughly. 
Compared with~Wu et al.~\cite{wu}, because of the distribution shift introduced by the directly-add synthesis approach, it hardly removes the nighttime flare and cannot remove the daytime flare thoroughly. 
Dai et al.~\cite{dai}  propose to specially remove nighttime flare thus performing worse in the daytime cases. 
With only \cite{wu}'s training set, our method exhibits better flare removal in both daytime and nighttime cases.

\noindent \textbf{Quantitative Evaluation:} 
We use  full-reference metrics PSNR and SSIM~\cite{ssim} to evaluate the performance of different methods. 
The scores in Table.~\ref{tab} are calculated on the test images provided in Flare 7K~\cite{dai} because this dataset has the paired data in both daytime and nighttime. 
Table.~\ref{tab} shows that the model trained by our synthesis pipeline and Wu et al. \cite{wu} dataset attains the best result under the model of U-Net. Our method achieves slight improvements when using Flare7k~\cite{dai} training set because all the flare images in \cite{dai} are synthetic. 
We also use a transform-based model Uformer~\cite{Uformer} to test our synthesis pipeline. It increases in PSNR but decreases in SSIM.

\noindent \textbf{User Study:}
We conduct a user study to compare our approach with~\cite{wu, he, traditional, zhang} trained under two datasets~\cite{wu, dai}. 
We use these five methods to produce flare-free images. 
Each time, participants are presented with two  flare-free images produced by two methods. They are asked to vote for which one has a better result. Table.~\ref{user} shows that more participants recognize the model trained using our method. The model using our method trained on \cite{wu} dataset improves a lot on the \cite{dai} test set and our consumer electronics testset.
%\lichongyi{what is our test set?}. 
We also train our model using Flare7K~\cite{dai} dataset. The performance of our approach has been consistently recognized. 
\subsection{Light Source Recovery Comparison}
 %\vspace{-.2cm}
\label{recover}

\begin{figure}[t]
  \begin{center}
  \begin{subfigure}{0.32\linewidth}
    \caption{Input}
    \includegraphics[width=2.65cm,height=2.65cm]{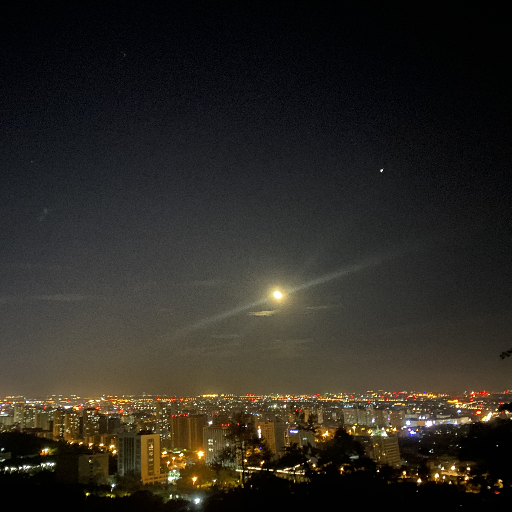}
  \end{subfigure}
  \begin{subfigure}{0.32\linewidth}
    \caption{~\cite{wu, dai}}
    \includegraphics[width=2.65cm,height=2.65cm]{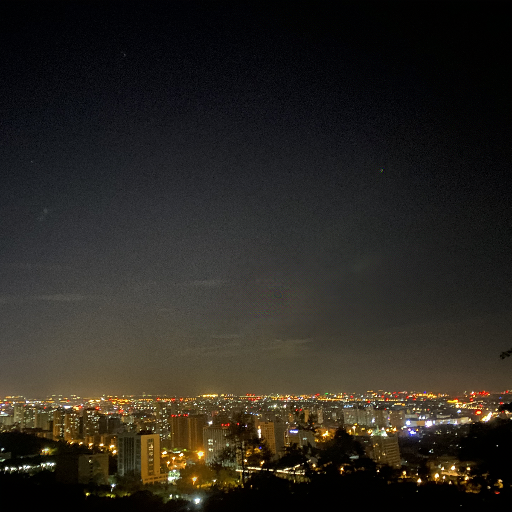}
  \end{subfigure}
  \begin{subfigure}{0.32\linewidth}
    \caption{Ours}
    \includegraphics[width=2.65cm,height=2.65cm]{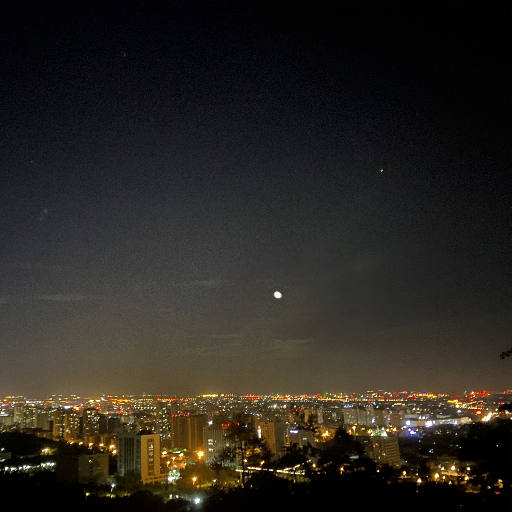}
  \end{subfigure}\\
  \begin{subfigure}{0.32\linewidth}
    \includegraphics[width=2.65cm,height=2.65cm]{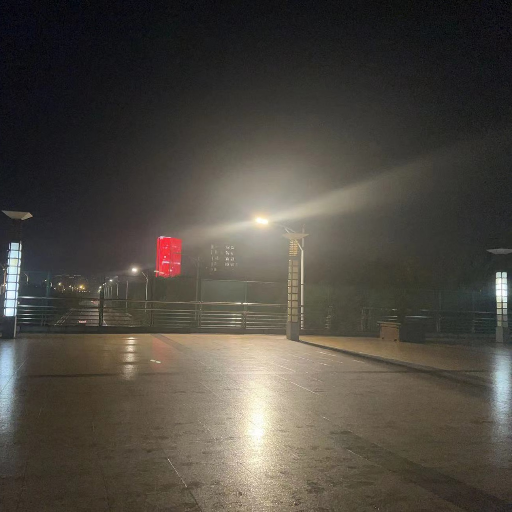}
  \end{subfigure}
  \begin{subfigure}{0.32\linewidth}
    \includegraphics[width=2.65cm,height=2.65cm]{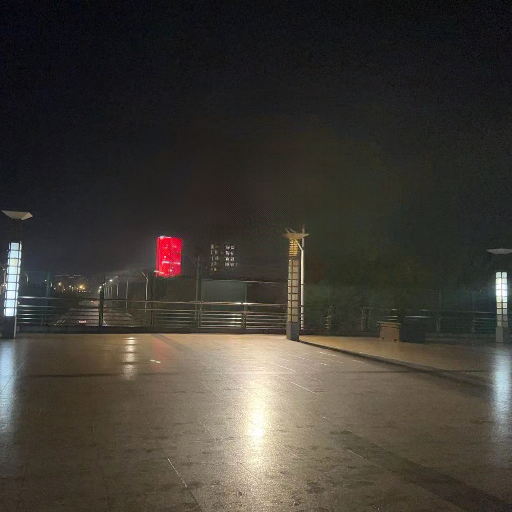}
  \end{subfigure}
  \begin{subfigure}{0.32\linewidth}
    \includegraphics[width=2.65cm,height=2.65cm]{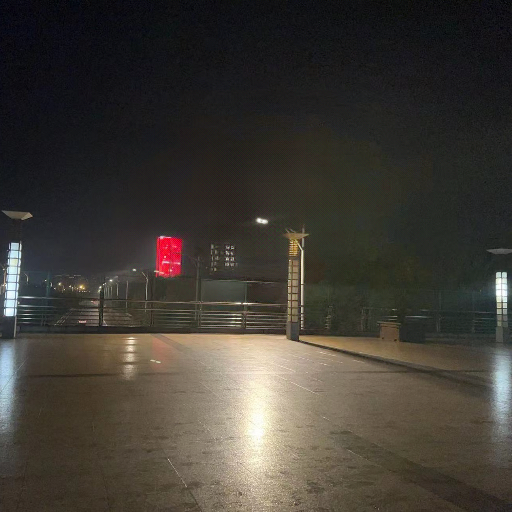}
  \end{subfigure}
  \end{center}
   %\vspace{-.41cm}  
  \caption{\small Single light source recovery on real images.
  }
  \label{single}
\end{figure}

\noindent \textbf{Single Light Source:}
In current methods~\cite{wu}, they first set a threshold such as $0.99$ to choose the candidates of the light source and apply a smoothing filter. 
It performs well when the light source is bright enough. However, if the light is not that bright, it cannot be recovered. 
As pointed out in \cite{dai}, most of the time, the light at nighttime will not be larger than $0.99$. 
Figure.~\ref{single} shows that our method can recover the moon and street lamp at nighttime while \cite{wu, dai} fails.

\begin{figure}[t]
  \begin{center}
  \begin{subfigure}{0.32\linewidth}
    \caption{Input}
    \includegraphics[width=2.65cm,height=2.65cm]{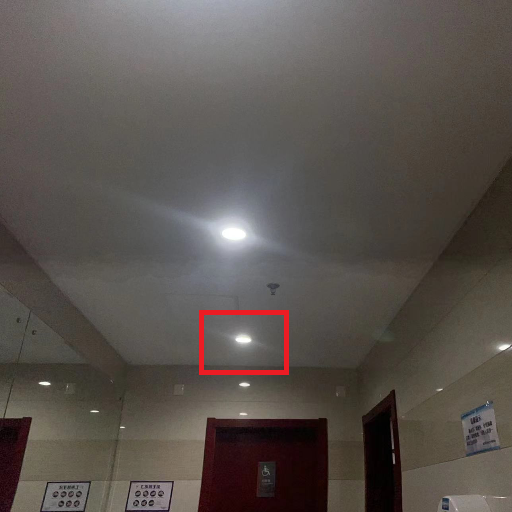}
  \end{subfigure}
  \begin{subfigure}{0.32\linewidth}
    \caption{~\cite{wu, dai}}
    \includegraphics[width=2.65cm,height=2.65cm]{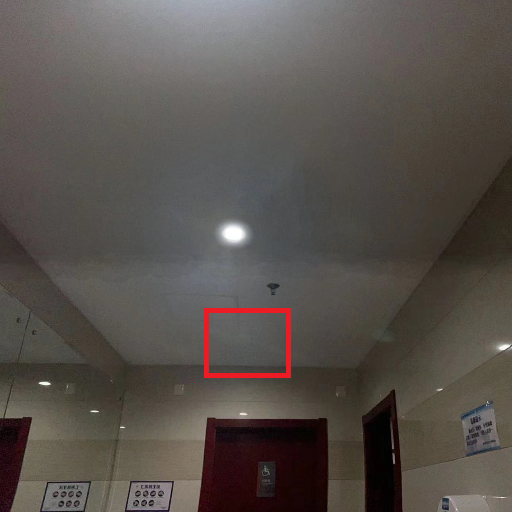}
  \end{subfigure}
  \begin{subfigure}{0.32\linewidth}
    \caption{Ours}
    \includegraphics[width=2.65cm,height=2.65cm]{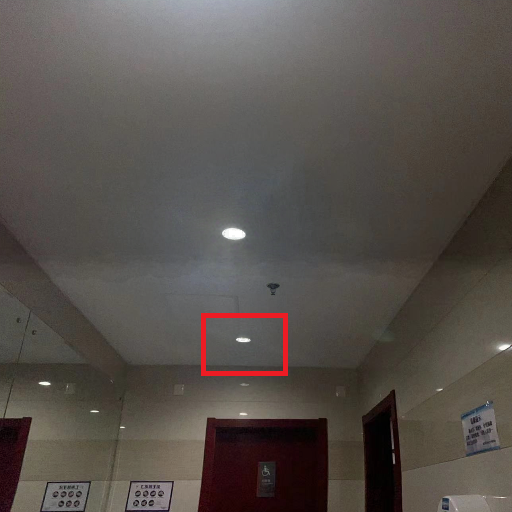}
  \end{subfigure}\\
  \begin{subfigure}{0.32\linewidth}
    \includegraphics[width=2.65cm,height=2.65cm]{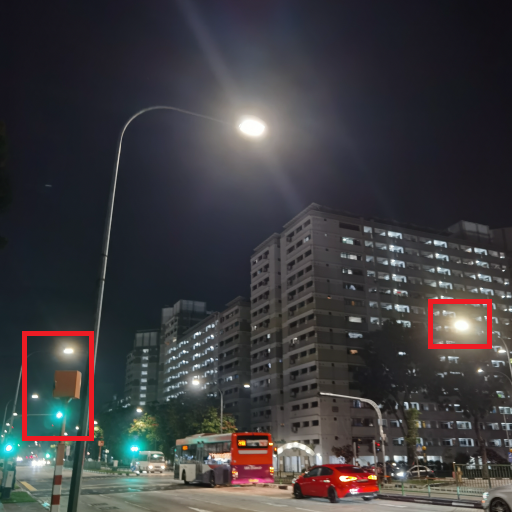}
  \end{subfigure}
  \begin{subfigure}{0.32\linewidth}
    \includegraphics[width=2.65cm,height=2.65cm]{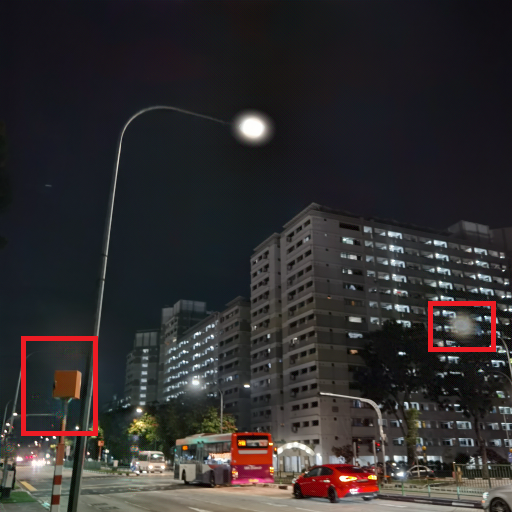}
  \end{subfigure}
  \begin{subfigure}{0.32\linewidth}
    \includegraphics[width=2.65cm,height=2.65cm]{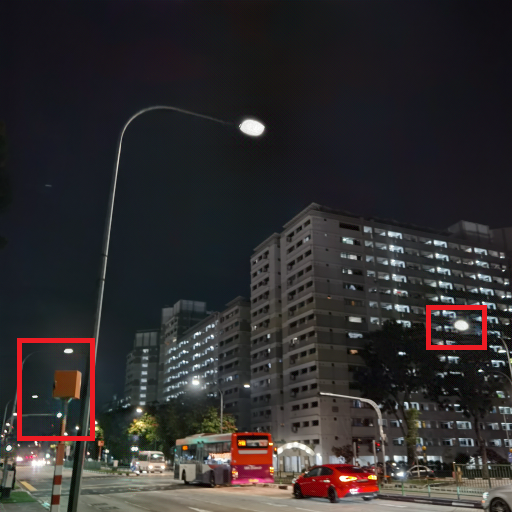}
  \end{subfigure}
  \end{center}
   %\vspace{-.4cm}
  \caption{\small Multiple light sources recovery on real images.} 
  %\vspace{-.7cm}
  \label{multiple}
\end{figure}

\noindent \textbf{Multiple Light Sources:} 
Figure.~\ref{multiple} shows that current light source recovery methods~\cite{wu, dai} can only recover the most conspicuous light source, but cannot recover the small light sources in the background. 
In contrast, our method recovers all the light sources with different sizes and positions well.

\begin{figure}[b]
    \centering
    %\vspace{-.3cm}
    \begin{subfigure}{0.23\linewidth}
    \includegraphics[width=\textwidth]{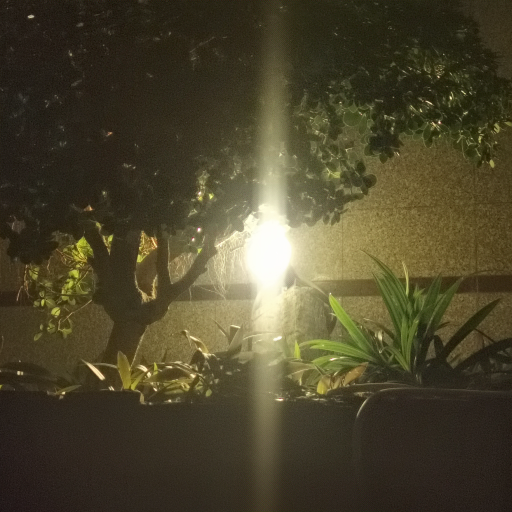}
    \caption*{{$\alpha$=1}}
    \end{subfigure}
    \begin{subfigure}{0.23\linewidth}
    \includegraphics[width=\textwidth]{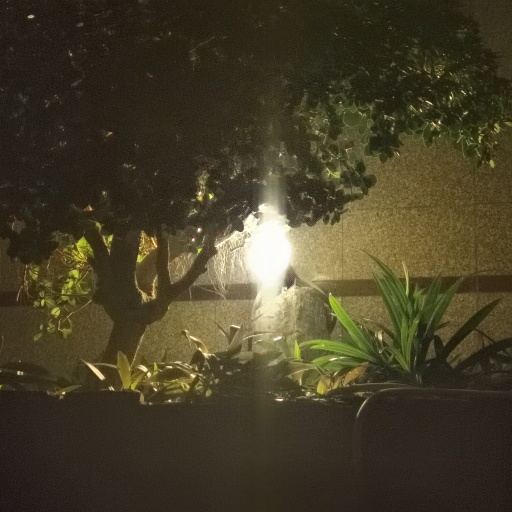}
    \caption*{{$\alpha$=5}}
    \end{subfigure}
    \begin{subfigure}{0.23\linewidth}
    \includegraphics[width=\textwidth]{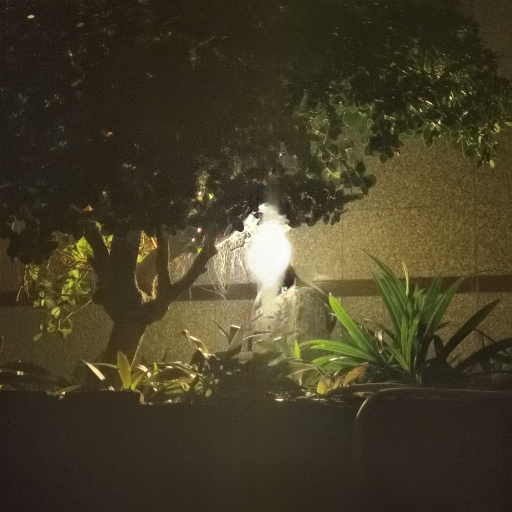}
    \caption*{{$\alpha$=15}}
    \end{subfigure}
    \begin{subfigure}{0.23\linewidth}
    \includegraphics[width=\textwidth]{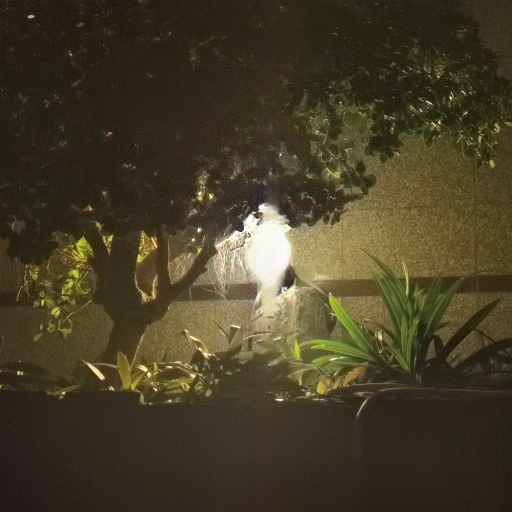}
    \caption*{{$\alpha$=20}}
    \end{subfigure}
    \caption{ Qualitative Comparison of different $\alpha$. 
    }
    \label{q2}
\end{figure}

\noindent \textbf{Comparison of different $\mathbf{\alpha}$:}
We compare different $\alpha$ on the performance of our light source recovery  qualitatively and quantitatively. 
Table.~\ref{q1} 
and Figure.~\ref{q2} 
show that when $\alpha>15$, PSNR and SSIM %\lichongyi{on which test set?}
maintain stability at 17.88 and 0.527. If $\alpha$ is too small, the flare will also be blended into the final image. Thus, we choose $\alpha=15$ as the default setting.   

\subsection{Generalization Comparison}
%\vspace{-.15cm}
As mentioned in Wu et al.~\cite{wu}, all reflective flares of dataset are captured with the same camera, distance, and focal length $f=13mm$. 
However, cameras of different smartphones have different focal lengths and the distance of light source varies a lot. Since the current flare removal test set only contains limited flare types, camera models, and light source types, it constrains the comparison of the generalization capability of different methods. 

\begin{figure*}[htb]
\begin{center}
  %input
  \begin{minipage}[b]{0.001\textwidth}
  {\rotatebox{90}{\footnotesize $\quad\quad$ Input}}
  \end{minipage}$\quad$
  \begin{subfigure}{0.13\linewidth}
    \caption*{\footnotesize iPhone 13 Pro}
    \includegraphics[height=2.2cm, width=2.3cm]{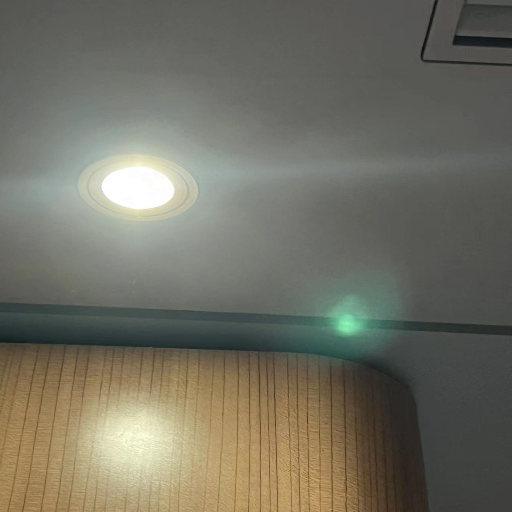}
  \end{subfigure}
  \begin{subfigure}{0.13\linewidth}
    \caption*{\footnotesize OPPO reno 4 Pro}
    \includegraphics[height=2.2cm, width=2.3cm]{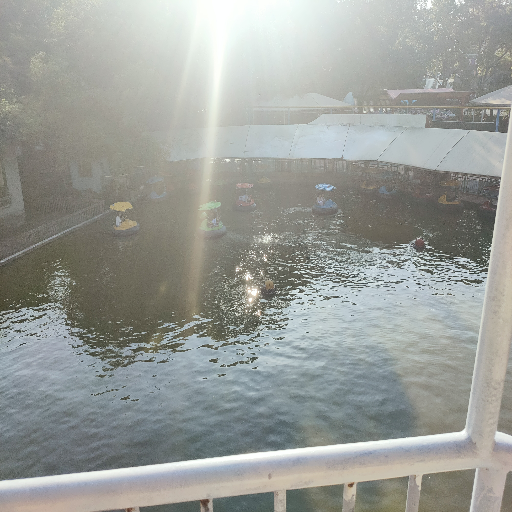}
  \end{subfigure}
  \begin{subfigure}{0.13\linewidth}
    \caption*{\footnotesize Huawei Mate 20}
    \includegraphics[height=2.2cm, width=2.3cm]{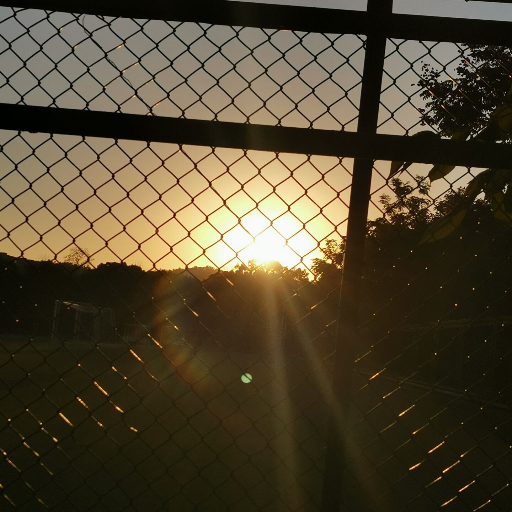}
  \end{subfigure}
  \begin{subfigure}{0.13\linewidth}
    \caption*{\footnotesize Xiaomi 12S Ultra}
    \includegraphics[height=2.2cm, width=2.3cm]{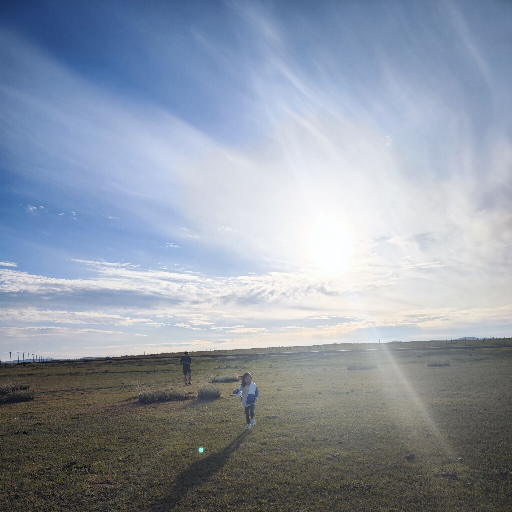}
  \end{subfigure}
  \begin{subfigure}{0.13\linewidth}
    \caption*{\footnotesize iPhone 11}
    \includegraphics[height=2.2cm, width=2.3cm]{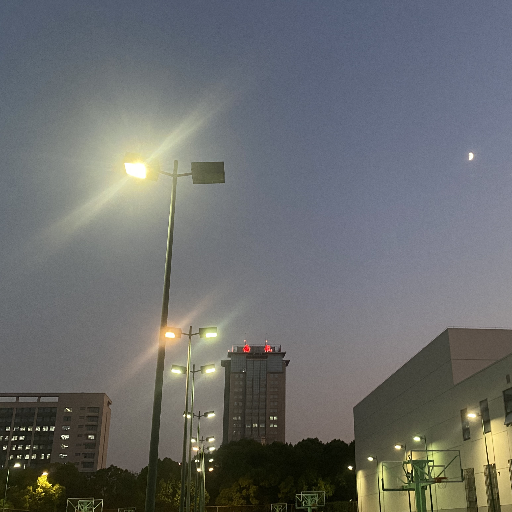}
  \end{subfigure}
  \begin{subfigure}{0.13\linewidth}
    \caption*{\footnotesize iPad Air4}
    \includegraphics[height=2.2cm, width=2.3cm]{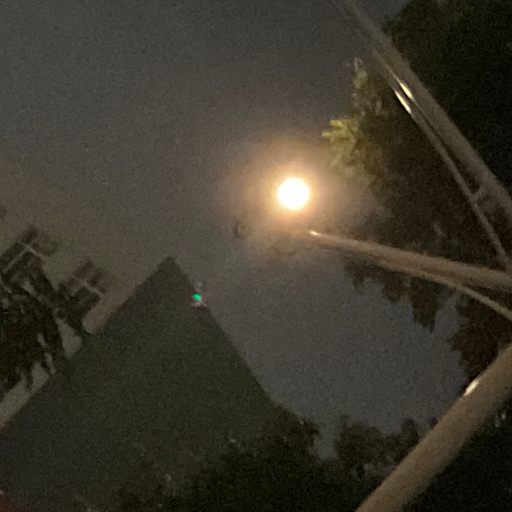}
  \end{subfigure}
    \begin{subfigure}{0.13\linewidth}
    \caption*{\footnotesize iPad 2020}
    \includegraphics[height=2.2cm, width=2.3cm]{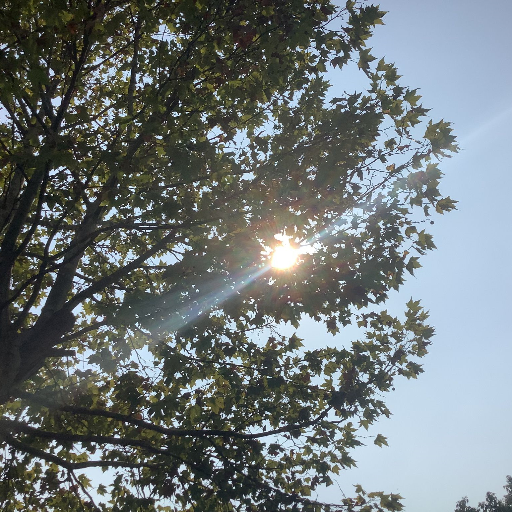}
  \end{subfigure}%\vspace{-1.3mm}
  \\
  %wu%%%%%%%%%%%%%%%%%%%%%%%%%%%%%%%%%%%%%%%%%%%%%%%%%%%%%%%%%%%%
  \begin{minipage}[b]{0.001\textwidth}
  {\rotatebox{90}{\footnotesize$\;\;\;$ Dai \cite{dai}+U-Net}}
  \end{minipage}$\quad$
  \begin{subfigure}{0.13\linewidth}
    \includegraphics[height=2.2cm, width=2.3cm]{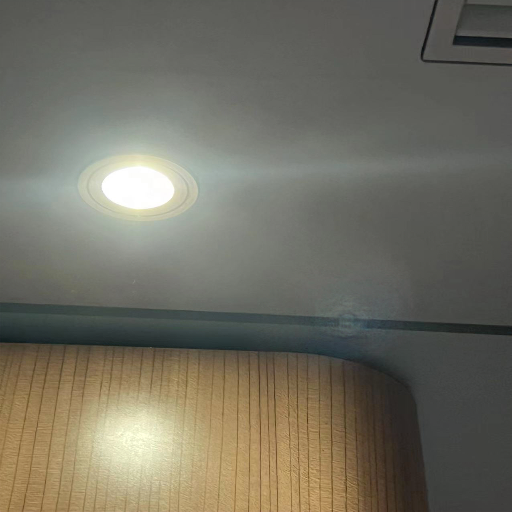}
  \end{subfigure}
  \begin{subfigure}{0.13\linewidth}
    \includegraphics[height=2.2cm, width=2.3cm]{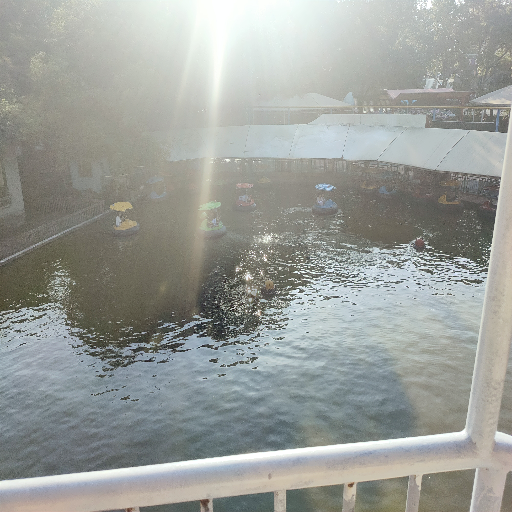}
  \end{subfigure}
  \begin{subfigure}{0.13\linewidth}
    \includegraphics[height=2.2cm, width=2.3cm]{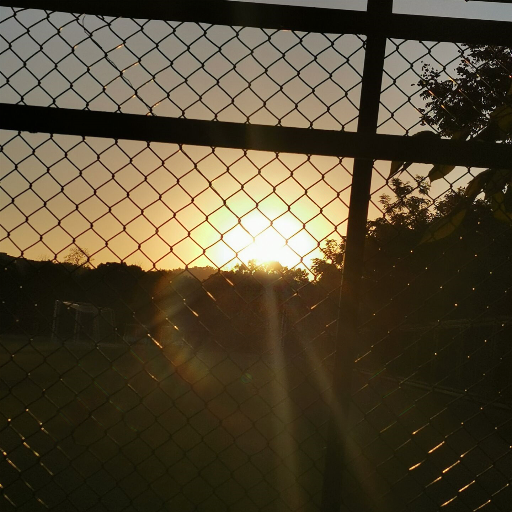}
  \end{subfigure}
  \begin{subfigure}{0.13\linewidth}
    \includegraphics[height=2.2cm, width=2.3cm]{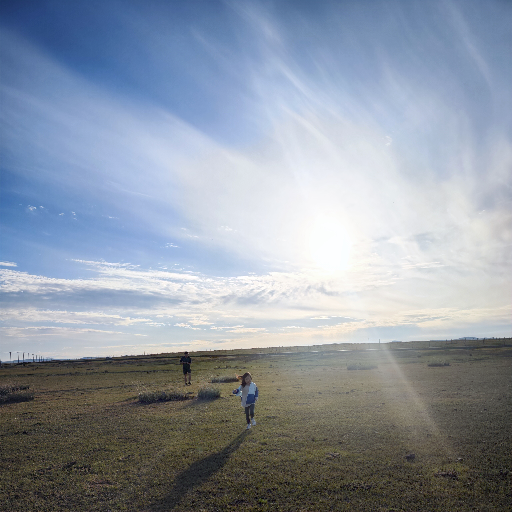}
  \end{subfigure}
  \begin{subfigure}{0.13\linewidth}
    \includegraphics[height=2.2cm, width=2.3cm]{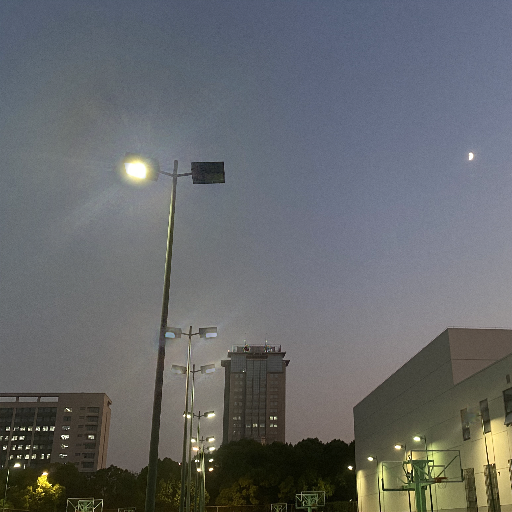}
  \end{subfigure}
  \begin{subfigure}{0.13\linewidth}
    \includegraphics[height=2.2cm, width=2.3cm]{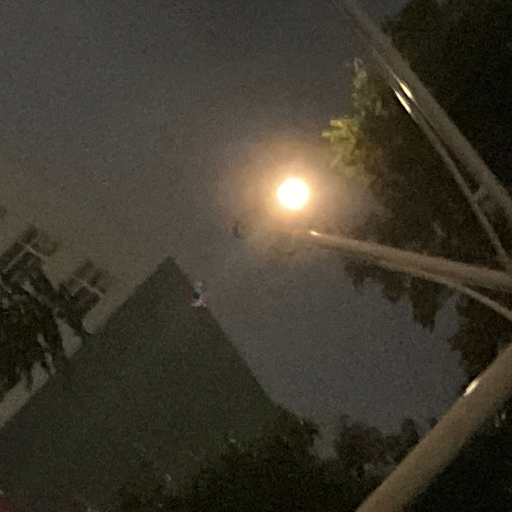}
  \end{subfigure}
   \begin{subfigure}{0.13\linewidth}
    \includegraphics[height=2.2cm, width=2.3cm]{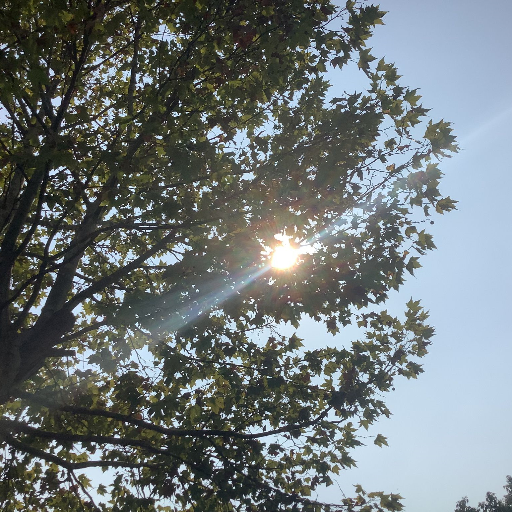}
  \end{subfigure}%\vspace{-1mm}
  \\
%dai%%%%%%%%%%%%%%%%%%%%%%%%%%%%%%%%%%%%%%%%%%%%%%%%%%%%%%%%%%%%%%%
  \begin{minipage}[b]{0.001\textwidth}
  {\rotatebox{90}{\footnotesize\;\; Wu \cite{wu}+U-Net}}
  \end{minipage}$\quad$
  \begin{subfigure}{0.13\linewidth}
    \includegraphics[height=2.2cm, width=2.3cm]{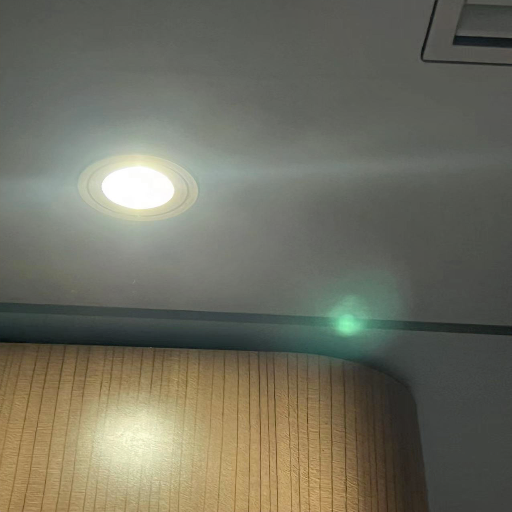}
  \end{subfigure}
  \begin{subfigure}{0.13\linewidth}
    \includegraphics[height=2.2cm, width=2.3cm]{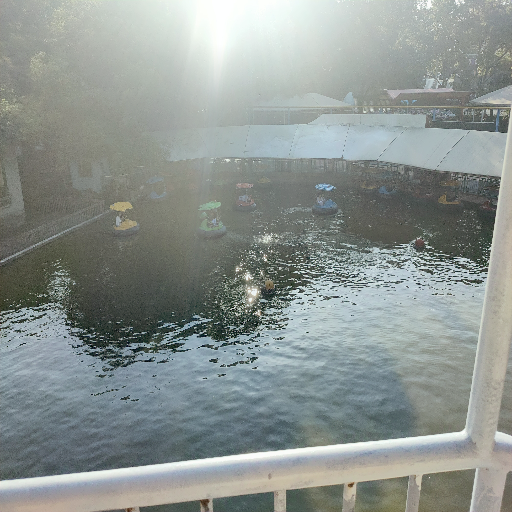}
  \end{subfigure}
  \begin{subfigure}{0.13\linewidth}
    \includegraphics[height=2.2cm, width=2.3cm]{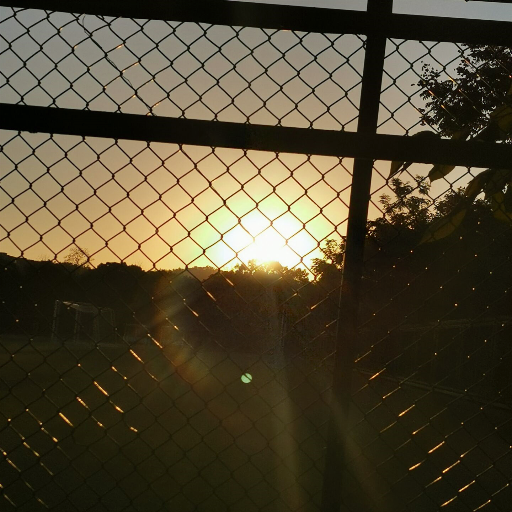}
  \end{subfigure}
  \begin{subfigure}{0.13\linewidth}
    \includegraphics[height=2.2cm, width=2.3cm]{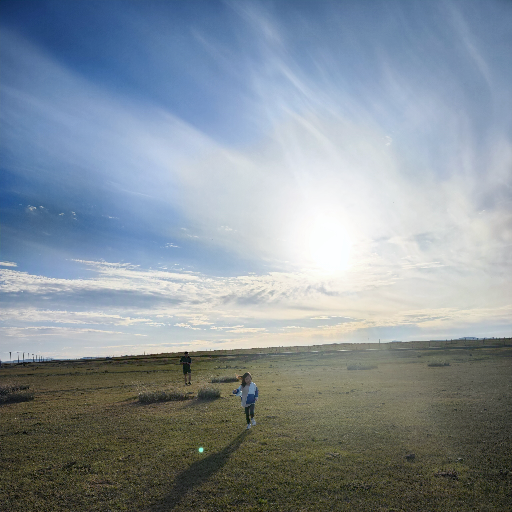}
  \end{subfigure}
  \begin{subfigure}{0.13\linewidth}
    \includegraphics[height=2.2cm, width=2.3cm]{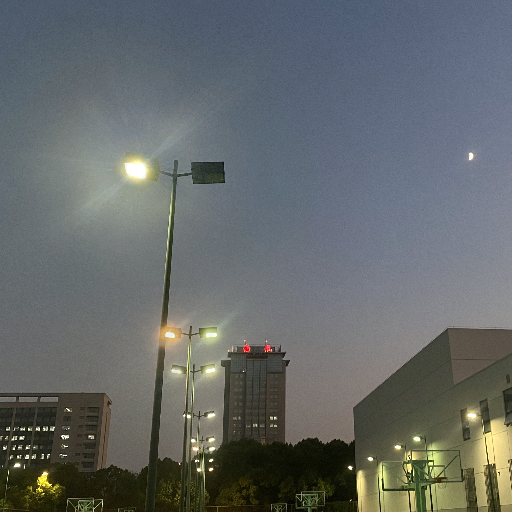}
  \end{subfigure}
  \begin{subfigure}{0.13\linewidth}
    \includegraphics[height=2.2cm, width=2.3cm]{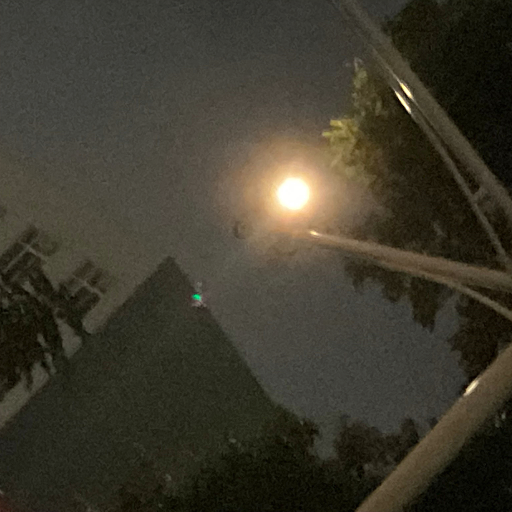}
  \end{subfigure}
  \begin{subfigure}{0.13\linewidth}
    \includegraphics[height=2.2cm, width=2.3cm]{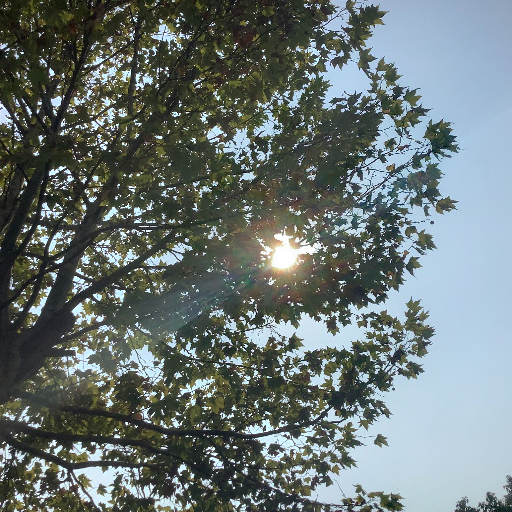}
  \end{subfigure}\\
%Ours+UNet%%%%%%%%%%%%%%%%%%%%%%%%%%%%%%%%%%%%%%%%%%%%%%%%%%%%%%%%%%%%
  \begin{minipage}[b]{0.001\textwidth}
  {\rotatebox{90}{\footnotesize$\;\;\;$ Ours+U-Net}}
  \end{minipage}$\quad$
  \begin{subfigure}{0.13\linewidth}
    \includegraphics[height=2.2cm, width=2.3cm]{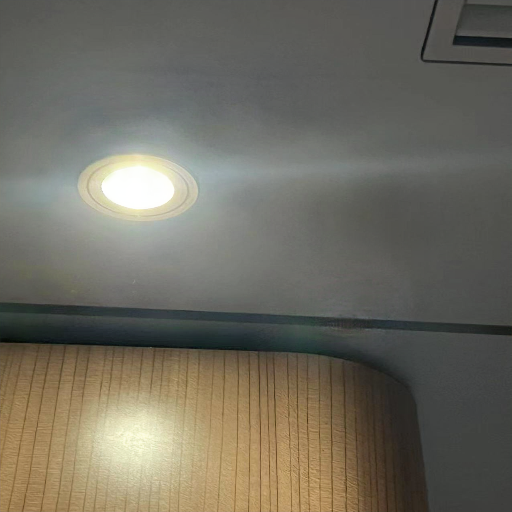}
  \end{subfigure}
  \begin{subfigure}{0.13\linewidth}
    \includegraphics[height=2.2cm, width=2.3cm]{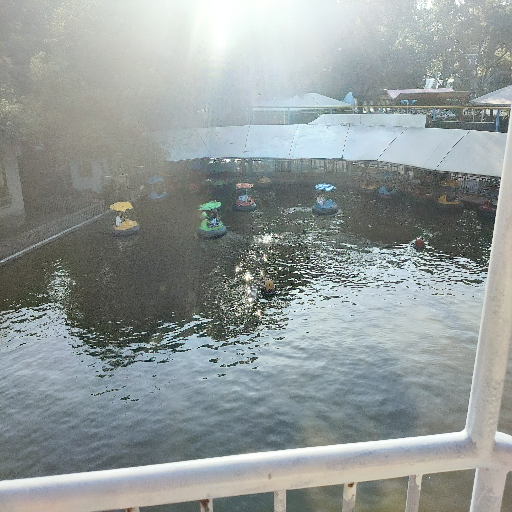}
  \end{subfigure}
  \begin{subfigure}{0.13\linewidth}
    \includegraphics[height=2.2cm, width=2.3cm]{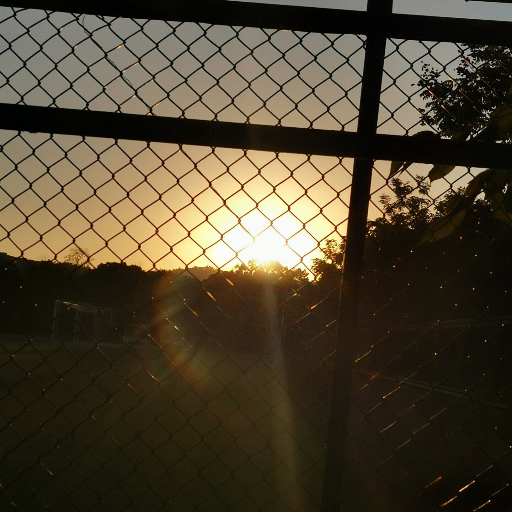}
  \end{subfigure}
  \begin{subfigure}{0.13\linewidth}
    \includegraphics[height=2.2cm, width=2.3cm]{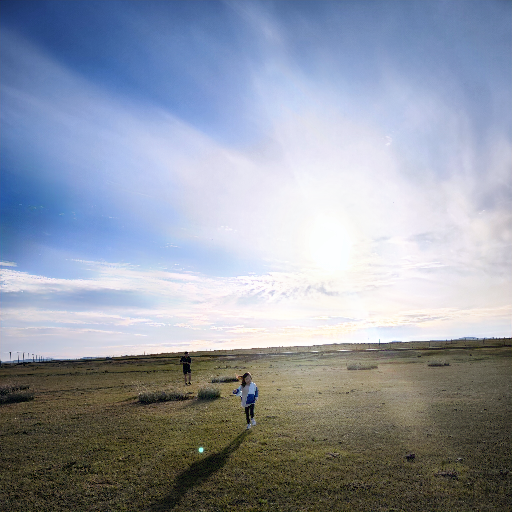}
  \end{subfigure}
  \begin{subfigure}{0.13\linewidth}
    \includegraphics[height=2.2cm, width=2.3cm]{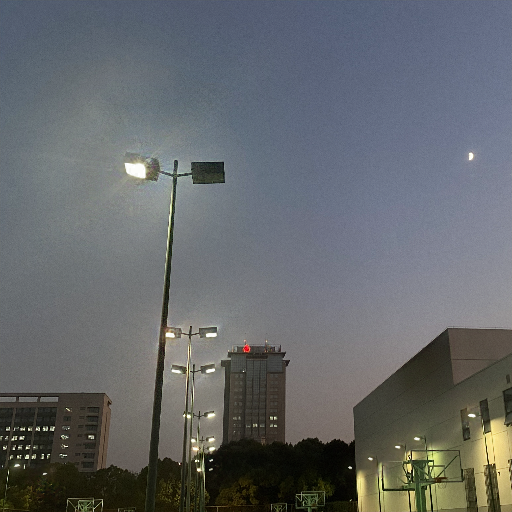}
  \end{subfigure}
  \begin{subfigure}{0.13\linewidth}
    \includegraphics[height=2.2cm, width=2.3cm]{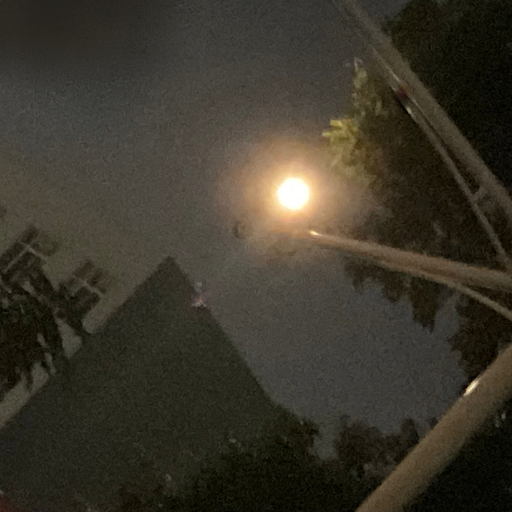}
  \end{subfigure}
  \begin{subfigure}{0.13\linewidth}
    \includegraphics[height=2.2cm, width=2.3cm]{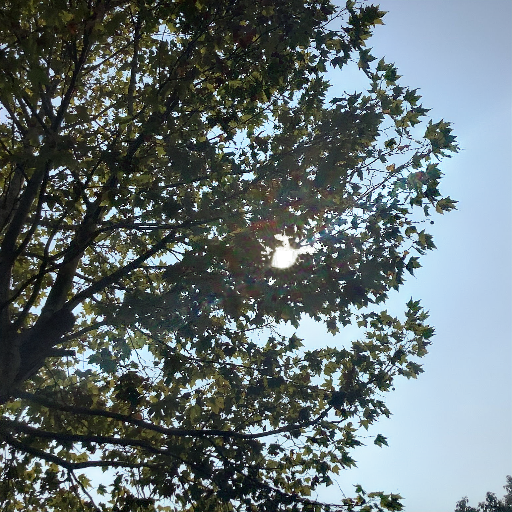}
  \end{subfigure}\\
  %Ours+Uformer
\begin{minipage}[b]{0.001\textwidth}
  {\rotatebox{90}{\footnotesize$\;$ Ours+Uformer}}
  \end{minipage}$\quad$
  \begin{subfigure}{0.13\linewidth}
    \includegraphics[height=2.2cm, width=2.3cm]{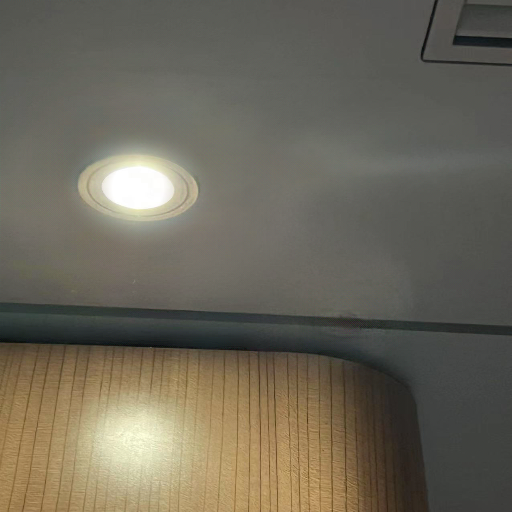}
  \end{subfigure}
  \begin{subfigure}{0.13\linewidth}
    \includegraphics[height=2.2cm, width=2.3cm]{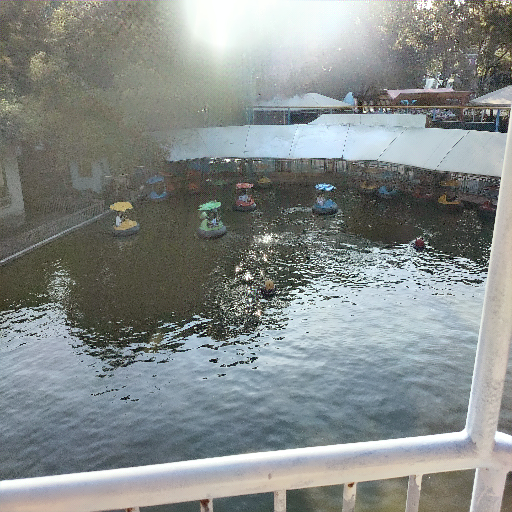}
  \end{subfigure}
  \begin{subfigure}{0.13\linewidth}
    \includegraphics[height=2.2cm, width=2.3cm]{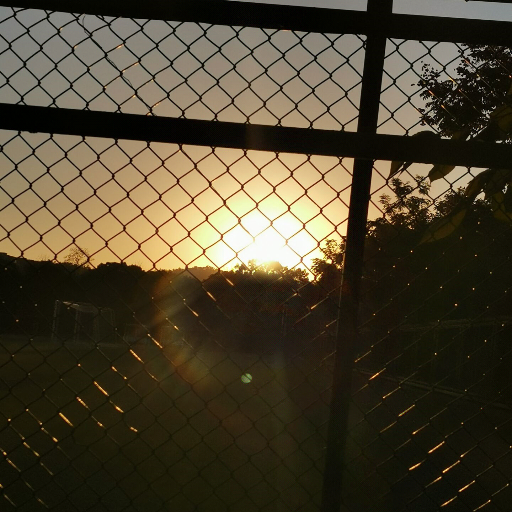}
  \end{subfigure}
  \begin{subfigure}{0.13\linewidth}
    \includegraphics[height=2.2cm, width=2.3cm]{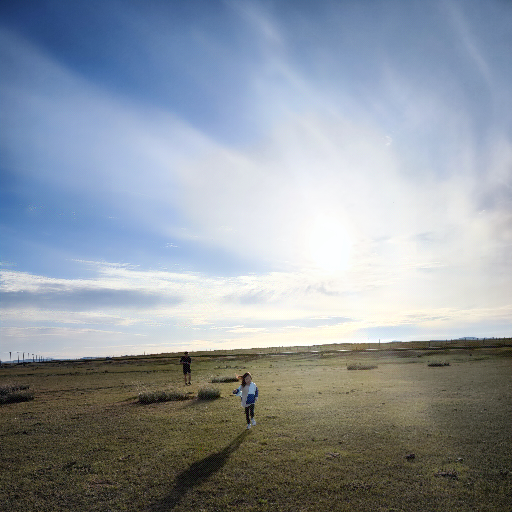}
  \end{subfigure}
  \begin{subfigure}{0.13\linewidth}
    \includegraphics[height=2.2cm, width=2.3cm]{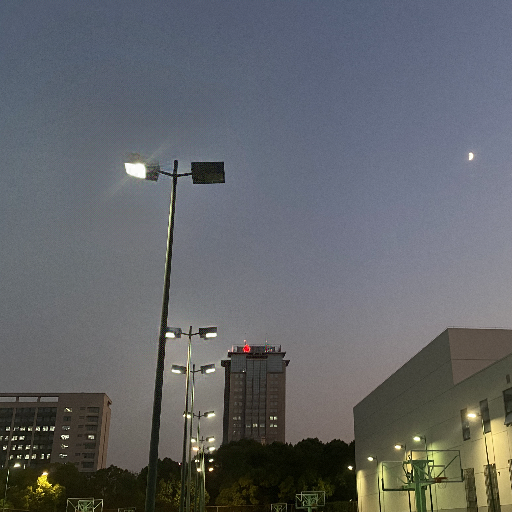}
  \end{subfigure}
  \begin{subfigure}{0.13\linewidth}
    \includegraphics[height=2.2cm, width=2.3cm]{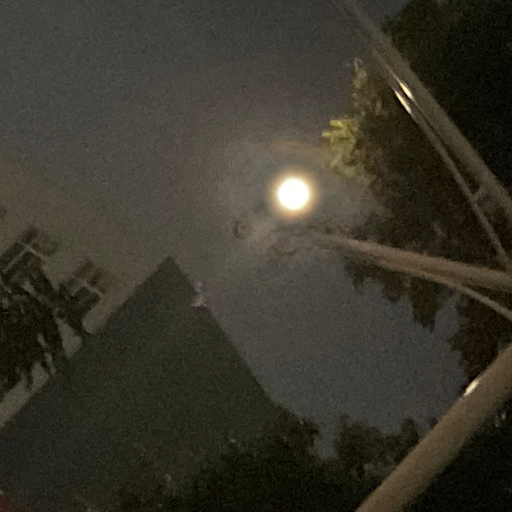}
  \end{subfigure}
  \begin{subfigure}{0.13\linewidth}
    \includegraphics[height=2.2cm, width=2.3cm]{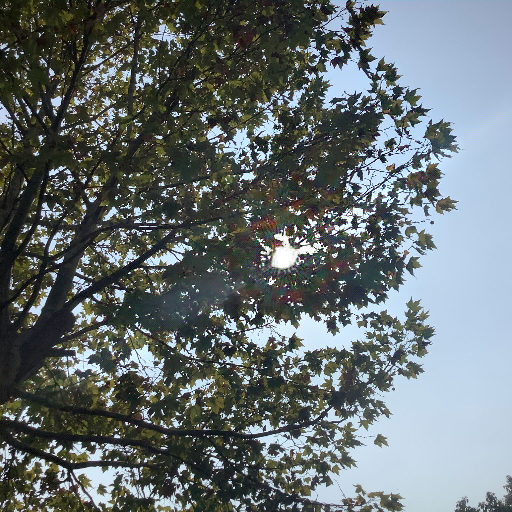}
  \end{subfigure}
  \caption{Visual comparison on our Consumer Electronics test dataset. 
  }
  %\vspace{-.75cm}
  \label{generalization}
\end{center}
\end{figure*}

To solve this issue, we collect an unpaired Consumer Electronics test dataset for evaluation. 
Flare images in our dataset are captured in both daytime and nighttime. 
For camera models, it contains 100 images captured by ten different cameras, including iPhone 13 pro, iPhone 11, Xiaomi 12S Ultra, Xiaomi 11,  iPad Air4, iPad 2020, Huawei Matepad, Vivo reno 4 pro, Huawei Mate 40 and Huawei Mate 20. 
%\lichongyi{please list all cameras.}.
%
For flare patterns, compared with Flare7K~\cite{dai} test set that only contains flare streak and flare haze and Wu et al.~\cite{wu} test set that only contains flare streak, flare blob, and color bleeding, our dataset contains richer flare shapes including streak, spot, blob, haze, and color bleeding. 
For light source types, the flares are taken under different light sources such as the sun, moon, street lamp, flashbulbs, etc.
Figure.~\ref{generalization} shows the generalization comparison of models trained using different synthesis methods. 
Our method can effectively remove different flares taken by different digital cameras. %(More results are provided in supplementary.)

%\vspace{-.15cm}
\subsection{Flare Removal for Object Detection}
\begin{figure}[htb]
  \centering
  %\scalebox{0.9}{
  \begin{subfigure}{0.24\linewidth}
    \includegraphics[scale=0.11]{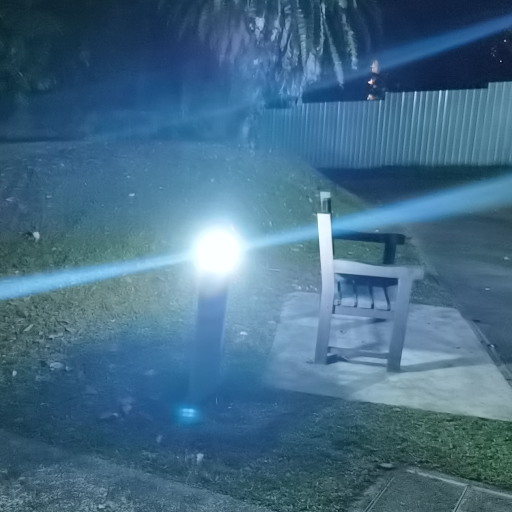}
    \label{id1}
  \end{subfigure}
  \begin{subfigure}{0.24\linewidth}
    \includegraphics[scale=0.11]{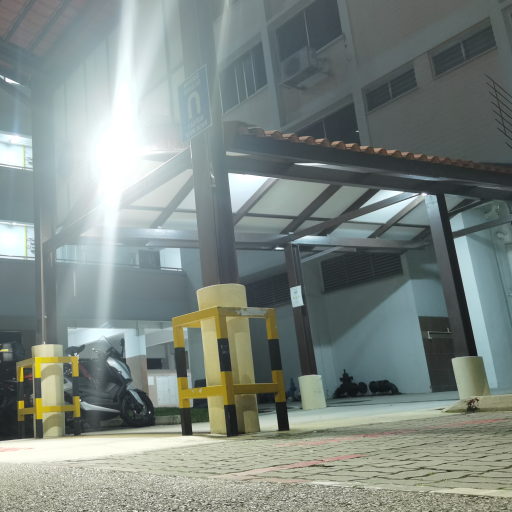}
    \label{id2}
  \end{subfigure}
  \begin{subfigure}{0.24\linewidth}
    \includegraphics[scale=0.11]{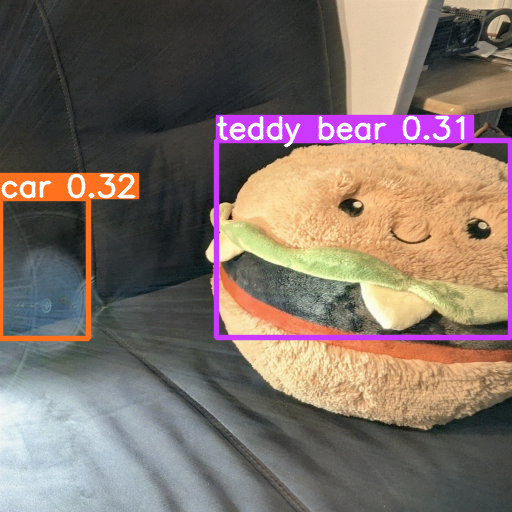}
    \label{id3}
  \end{subfigure}
  \begin{subfigure}{0.24\linewidth}
    \includegraphics[scale=0.11]{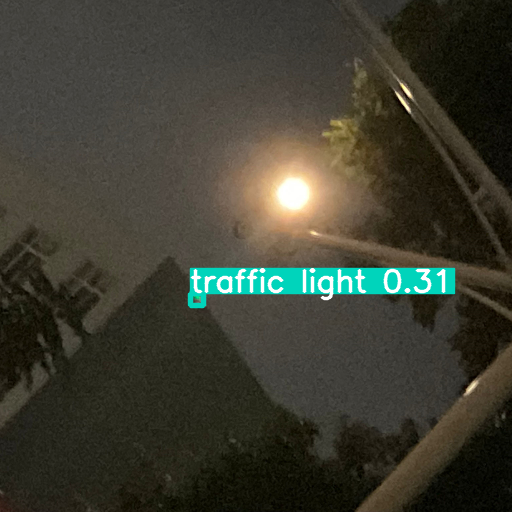}
    \label{id4}
  \end{subfigure}\\
    %\vspace{-.3cm}
  \begin{subfigure}{0.24\linewidth}
    \includegraphics[scale=0.11]{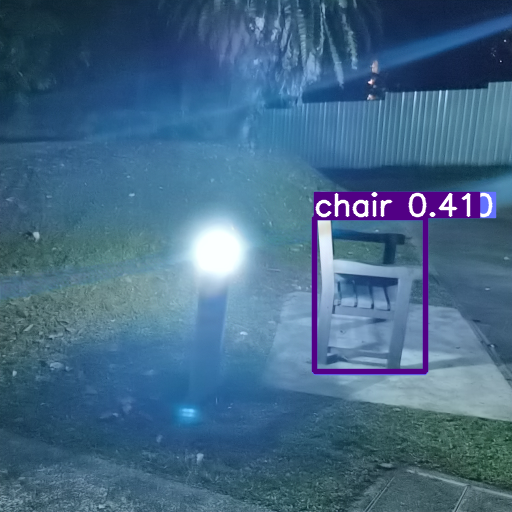}
  \end{subfigure}
  \begin{subfigure}{0.24\linewidth}
    \includegraphics[scale=0.11]{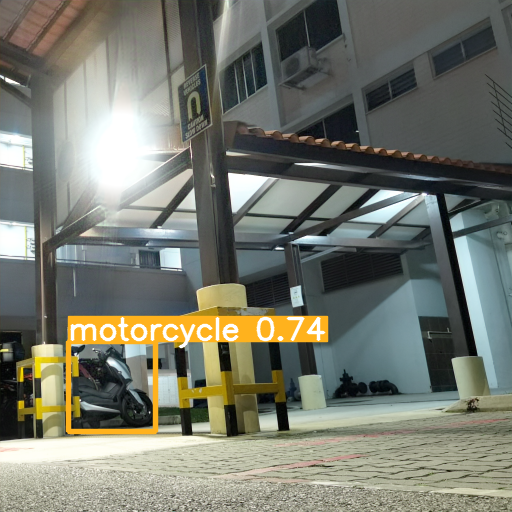}
  \end{subfigure}
  \begin{subfigure}{0.24\linewidth}
    \includegraphics[scale=0.11]{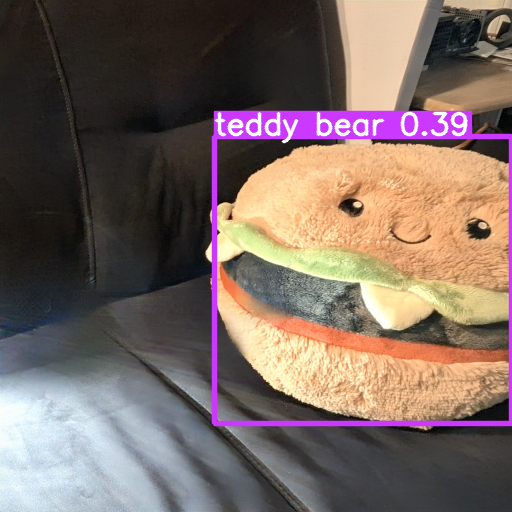}
  \end{subfigure}
  \begin{subfigure}{0.24\linewidth}
    \includegraphics[scale=0.11]{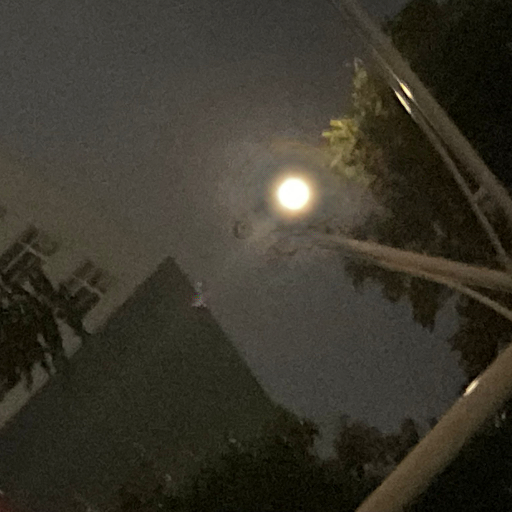}
  \end{subfigure}
  
  \caption{\small Object detection with flare (top) and after flare removal with the proposed solution (bottom).}
   \label{object detection}
\end{figure}

%\vspace{-.15cm}
Both scattering and reflective flares can pollute the images. 
To examine the influence of flare removal on object detection, we use pre-trained YOLOv5 detector to process the images with flares and flare removal results. 
For streak flare, it shades the image details so that detector cannot find the object. 
The first and second column in Figure.~\ref{object detection} shows that the flare streak shaded the chair and motorcycle, so the detector cannot detect it. 
%
%The second column demonstrates that the flare streak shades the  in the park. 
%
For reflective flare, the detector misunderstands it as an irrelevant object. 
The third and fourth columns show that the detector misunderstands the flare as a car and a traffic light. 
With our method to remove flare, the detector works better.
%\subsection{Limitation}
%\begin{itemize}
%  \item When the light source is bright enough and the flare exhibits like flare haze, our method will always failure.
%  \item Sometime when flare is removed, the background details occluded by flare cannot be recovered naturally.
%\end{itemize}
%\vspace{-.3cm}
\begin{table}[b]
%\begin{center}
%\vspace{-.4cm}
\centering
\caption{ Quantitative Comparison of different $\alpha$.}
\label{q1}
%\vspace{-.23cm}
\scalebox{.92}{
\begin{tabular}{|l|l|l|l|l|l|l|}
\hline
$\alpha$ & 1 & 5 & 10 & 15 & 20 & 25 \\ \hline
PSNR     & 15.94 & 17.68 & 17.86 & \textbf{17.88} & \textbf{17.88} & \textbf{17.88}     \\ \hline
SSIM     & 0.508 & 0.527 & \textbf{0.528} & \textbf{0.528} & \textbf{0.528} &  \textbf{0.528} \\ \hline
\end{tabular}
%\end{center}
}
\end{table}

\section{Conclusion}
In this paper, we proposed a new method to synthesize flare-corrupted images. 
Taking tone mapping into consideration, the flare-corrupted images synthesized using our method avoid distribution shift and overflow, making the flare removal model performs well. 
We also proposed a new method to smoothly recover multiple light sources. It uses a power function to soften the extraction range of the light source and avoid the hard threshold in other methods. 
To examine the generalization performance of flare removal methods, we contribute a new dataset that contains real flare-corrupted images captured by diverse consumer electronics for evaluation.
Extensive experiments show that the model trained using paired data synthesized by our practice can better remove lens flare, and our approach can recover multiple light sources effectively.  
\section{Acknowledgement}
We appreciate Tian Lan, Jiawei Liu, Shaoming Yan and Liang Zhu from NUAA for helping us collect the test dataset. 
This work is supported in part by the National Natural Science Foundation of China under grant 62272229, and the Natural Science Foundation of Jiangsu Province under grant BK20222012.
We also gratefully acknowledge the support of MindSpore, CANN, and Ascend AI Processor used for this research.

\newpage
%%%%%%%%% REFERENCES
{\small
\bibliographystyle{ieee_fullname}
\bibliography{egbib}

\begin{thebibliography}{10}\itemsep=-1pt

\bibitem{traditional}
CS Asha, Sooraj~Kumar Bhat, Deepa Nayak, and Chaithra Bhat.
\newblock Auto removal of bright spot from images captured against flashing
  light source.
\newblock In {\em 2019 IEEE International Conference on Distributed Computing,
  VLSI, Electrical Circuits and Robotics (DISCOVER)}, pages 1--6. IEEE, 2019.

\bibitem{unprocessing}
Tim Brooks, Ben Mildenhall, Tianfan Xue, Jiawen Chen, Dillon Sharlet, and
  Jonathan~T Barron.
\newblock Unprocessing images for learned raw denoising.
\newblock In {\em CVPR}, pages 11036--11045, 2019.

\bibitem{traditional2}
Floris Chabert.
\newblock Automated lens flare removal.
\newblock In {\em Technical report}. Department of Electrical Engineering,
  Stanford University, 2015.

\bibitem{antireflection}
Hou-Tong Chen, Jiangfeng Zhou, John~F O’Hara, Frank Chen, Abul~K Azad, and
  Antoinette~J Taylor.
\newblock Antireflection coating using metamaterials and identification of its
  mechanism.
\newblock {\em Physical review letters}, 105(7):073901, 2010.

\bibitem{hazeremoval1}
Wei-Ting Chen, Jian-Jiun Ding, and Sy-Yen Kuo.
\newblock Pms-net: Robust haze removal based on patch map for single images.
\newblock In {\em CVPR}, pages 11681--11689, 2019.

\bibitem{dai}
Yuekun Dai, Chongyi Li, Shangchen Zhou, Ruicheng Feng, and Chen~Change Loy.
\newblock Flare7k: A phenomenological nighttime flare removal dataset.
\newblock In {\em Thirty-sixth Conference on Neural Information Processing
  Systems Datasets and Benchmarks Track}, 2022.

\bibitem{hazeremoval2}
Akshay Dudhane and Subrahmanyam Murala.
\newblock Ryf-net: Deep fusion network for single image haze removal.
\newblock {\em IEEE Transactions on Image Processing}, 29:628--640, 2019.

\bibitem{hazeremoval4}
Akshay Dudhane, Harshjeet Singh~Aulakh, and Subrahmanyam Murala.
\newblock Ri-gan: An end-to-end network for single image haze removal.
\newblock In {\em CVPRW}, 2019.

\bibitem{fan}
Qingnan Fan, Jiaolong Yang, Gang Hua, Baoquan Chen, and David Wipf.
\newblock A generic deep architecture for single image reflection removal and
  image smoothing.
\newblock In {\em ICCV}, pages 3238--3247, 2017.

\bibitem{cenet}
Qingnan Fan, Jiaolong Yang, Gang Hua, Baoquan Chen, and David Wipf.
\newblock A generic deep architecture for single image reflection removal and
  image smoothing.
\newblock In {\em ICCV}, pages 3238--3247, 2017.

\bibitem{zerodce}
Chunle Guo, Chongyi Li, Jichang Guo, Chen~Change Loy, Junhui Hou, Sam Kwong,
  and Runmin Cong.
\newblock Zero-reference deep curve estimation for low-light image enhancement.
\newblock In {\em CVPR}, pages 1780--1789, 2020.

\bibitem{guo2022image}
Chun-Le Guo, Qixin Yan, Saeed Anwar, Runmin Cong, Wenqi Ren, and Chongyi Li.
\newblock Image dehazing transformer with transmission-aware 3d position
  embedding.
\newblock In {\em CVPR}, pages 5812--5820, 2022.

\bibitem{he}
Kaiming He, Jian Sun, and Xiaoou Tang.
\newblock Single image haze removal using dark channel prior.
\newblock {\em IEEE Transactions on Pattern Analysis and Machine intelligence},
  33(12):2341--2353, 2010.

\bibitem{rainremoval1}
Xiaowei Hu, Chi-Wing Fu, Lei Zhu, and Pheng-Ann Heng.
\newblock Depth-attentional features for single-image rain removal.
\newblock In {\em CVPR}, pages 8022--8031, 2019.

\bibitem{jin2023enhancing}
Yeying Jin, Beibei Lin, Wending Yan, Wei Ye, Yuan Yuan, and Robby~T Tan.
\newblock Enhancing visibility in nighttime haze images using guided apsf and
  gradient adaptive convolution.
\newblock {\em arXiv preprint arXiv:2308.01738}, 2023.

\bibitem{jin2022unsupervised}
Yeying Jin, Wenhan Yang, and Robby~T Tan.
\newblock Unsupervised night image enhancement: When layer decomposition meets
  light-effects suppression.
\newblock In {\em ECCV}, pages 404--421. Springer, 2022.

\bibitem{reflection3}
Chenyang Lei and Qifeng Chen.
\newblock Robust reflection removal with reflection-free flash-only cues.
\newblock In {\em CVPR}, pages 14811--14820, 2021.

\bibitem{reflection2}
Chenyang Lei, Xuhua Huang, Mengdi Zhang, Qiong Yan, Wenxiu Sun, and Qifeng
  Chen.
\newblock Polarized reflection removal with perfect alignment in the wild.
\newblock In {\em CVPR}, pages 1750--1758, 2020.

\bibitem{li2020single}
Chao Li, Yixiao Yang, Kun He, Stephen Lin, and John~E Hopcroft.
\newblock Single image reflection removal through cascaded refinement.
\newblock In {\em CVPR}, pages 3565--3574, 2020.

\bibitem{scl}
Dong Liang, Ling Li, Mingqiang Wei, Shuo Yang, Liyan Zhang, Wenhan Yang, Yun
  Du, and Huiyu Zhou.
\newblock Semantically contrastive learning for low-light image enhancement.
\newblock In {\em AAAI Conference on Artificial Intelligence}, volume~36, pages
  1555--1563, 2022.

\bibitem{liu}
Shu-yun Liu, Qun Hao, Yu-tong Zhang, Feng Gao, Hai-ping Song, Yu-tong Jiang,
  Ying-sheng Wang, Xiao-ying Cui, and Kun Gao.
\newblock Single-image night haze removal based on color channel transfer and
  estimation of spatial variation in atmospheric light.
\newblock {\em Defence Technology}, 2022.

\bibitem{tmoreview}
Ziyi Liu.
\newblock A review for tone-mapping operators on wide dynamic range image.
\newblock {\em arXiv preprint arXiv:2101.03003}, 2021.

\bibitem{qiao}
Xiaotian Qiao, Gerhard~P Hancke, and Rynson~WH Lau.
\newblock Light source guided single-image flare removal from unpaired data.
\newblock In {\em ICCV}, pages 4177--4185, 2021.

\bibitem{UNet}
Olaf Ronneberger, Philipp Fischer, and Thomas Brox.
\newblock U-net: Convolutional networks for biomedical image segmentation.
\newblock In {\em MICCAI}, pages 234--241. Springer, 2015.

\bibitem{traditional3}
Patricia Vitoria and Coloma Ballester.
\newblock Automatic flare spot artifact detection and removal in photographs.
\newblock {\em Journal of Mathematical Imaging and Vision}, 61(4):515--533,
  2019.

\bibitem{rainremoval3}
Cong Wang, Xiaoying Xing, Yutong Wu, Zhixun Su, and Junyang Chen.
\newblock Dcsfn: Deep cross-scale fusion network for single image rain removal.
\newblock In {\em ACMMM}, pages 1643--1651, 2020.

\bibitem{rainremoval2}
Hong Wang, Qi Xie, Qian Zhao, and Deyu Meng.
\newblock A model-driven deep neural network for single image rain removal.
\newblock In {\em CVPR}, pages 3103--3112, 2020.

\bibitem{ssim}
Zhou Wang, Alan~C Bovik, Hamid~R Sheikh, and Eero~P Simoncelli.
\newblock Image quality assessment: from error visibility to structural
  similarity.
\newblock {\em IEEE Transactions on Image Processing}, 13(4):600--612, 2004.

\bibitem{Uformer}
Zhendong Wang, Xiaodong Cun, Jianmin Bao, Wengang Zhou, Jianzhuang Liu, and
  Houqiang Li.
\newblock Uformer: A general u-shaped transformer for image restoration.
\newblock In {\em CVPR}, pages 17683--17693, 2022.

\bibitem{distribution}
Olivia Wiles, Sven Gowal, Florian Stimberg, Sylvestre Alvise-Rebuffi, Ira
  Ktena, Taylan Cemgil, et~al.
\newblock A fine-grained analysis on distribution shift.
\newblock {\em arXiv preprint arXiv:2110.11328}, 2021.

\bibitem{wu}
Yicheng Wu, Qiurui He, Tianfan Xue, Rahul Garg, Jiawen Chen, Ashok
  Veeraraghavan, and Jonathan~T Barron.
\newblock How to train neural networks for flare removal.
\newblock In {\em ICCV}, pages 2239--2247, 2021.

\bibitem{reflection1}
Jie Yang, Dong Gong, Lingqiao Liu, and Qinfeng Shi.
\newblock Seeing deeply and bidirectionally: A deep learning approach for
  single image reflection removal.
\newblock In {\em ECCV}, pages 654--669, 2018.

\bibitem{hazeremoval3}
Shengdong Zhang, Fazhi He, and Wenqi Ren.
\newblock Nldn: Non-local dehazing network for dense haze removal.
\newblock {\em Neurocomputing}, 410:363--373, 2020.

\bibitem{zhang}
Xuaner Zhang, Ren Ng, and Qifeng Chen.
\newblock Single image reflection separation with perceptual losses.
\newblock In {\em CVPR}, pages 4786--4794, 2018.

\end{thebibliography}
}

\end{document}